\documentclass[preprint]{revtex4}
\usepackage{graphicx,subfigure}
\usepackage{epsfig}

\begin{document}

\title{On the radial expansion of tubular structures  in a  quark gluon plasma}

\author{D.A. Foga\c{c}a\dag\,  F.S. Navarra\dag\ and L.G. Ferreira Filho\ddag\ }
\address{\dag\ Instituto de F\'{\i}sica, Universidade de S\~{a}o Paulo\\
 C.P. 66318,  05315-970 S\~{a}o Paulo, SP, Brazil}
\address{\ddag\ Faculdade de Tecnologia, Universidade do Estado do Rio de Janeiro \\
Via Dutra km 298, CEP 27523-000, Resende, RJ, Brazil}

\begin{abstract}

We study the radial expansion of cylindrical  tubes in a hot QGP. These tubes are  treated as perturbations in
the energy  density of the system which is formed in heavy ion collisions at RHIC and LHC. We start from the
equations of relativistic hydrodynamics in two spatial dimensions and cylindrical symmetry and perform an expansion
of these equations in a small parameter, conserving the nonlinearity of the hydrodynamical formalism. We consider both
ideal and viscous fluids and  the latter are studied with a relativistic Navier-Stokes equation. We use the
equation of state of  the MIT bag model. In the case of ideal fluids we obtain a breaking wave equation for the energy
density fluctuation, which is then solved numerically.  We also show that, under certain assumptions, perturbations in a
relativistic viscous fluid are governed by the Burgers equation.  We estimate the typical  expansion time of the tubes.

\end{abstract}

\maketitle



\vspace{1cm}
\section{Introduction}

The study of the initial stage of relativistic heavy ion collisions has experienced
a fast progress in recent years. One of the most interesting findings in this study,
supported both by theoretical works and by the analysis of experimental data, is that
in the early times of these collisions  color flux tubes are formed. Although color
flux tubes are familiar objects in hadron physics it is not obvious that they should
be formed in high energy heavy ion collisions, where projectile and target can be
regarded as bunches of partons without any strong clustering neither in configuration
nor in color space. Flux tubes, sometimes called strings, appear in lattice QCD calculations
as field configurations between static heavy charges. They appear also in phenomenological
models of high energy soft hadronic scattering. In the Lund  model, for example, when two
high energy protons collide with low momentum transfer, they cross each other and, due to
gluon exchange, a color rearrangement takes place with the subsequent formation of strings,
which stretch and decay, producing  particles.

At high energies, colliding nuclei are
dense systems in which the standard linear evolution equations (such as the DGLAP equations)
must be replaced by others, which include nonlinear effects in the evolution. The theoretical
description of these dense systems evolved into the  theory of the Color Glass Condensate
(CGC). In this formalism the dense gluonic matter is treated in a semi-classical approximation.
In this approach, it has been shown in  \cite{cgc1,cgc2,cgc3,cgc4} that there are solutions of the
classical Yang-Mills equations in which the lines of the chromo-electric and chromo-magnetic fields
are all parallel to the collision axis and these fields form color flux tubes in the longitudinal
(z) direction.

The interpretation of RHIC and LHC data also suggests that the system reaches thermal equilibrium, forming
a thermalized quark-gluon plasma (QGP), very soon after the collision. At first sight this would imply that
the flux tubes disappear and the quark-gluon matter becomes reasonably homogeneous when the hydrodynamical
expansion starts. However detailed hydrodynamical studies  \cite{hamatube,andrade} strongly suggest
that some experimental features observed at RHIC and LHC can be understood  if we assume
that these tubes survive the thermalization stage and form ``tubular'' structures that persist for some
time during the  hydrodynamical expansion. More specifically, the data show
the existence of structures in the two-particle correlations plotted as function of the pseudorapidity difference
$\Delta \eta$ and  the angular spacing $\Delta \phi$. In \cite{hamatube,andrade} it has been argued that
these structures may have a common hydrodynamic origin: the combined effect of longitudinal high energy
density tubes (leftover from initial particle collisions) and transverse expansion.

The tubular structures described above, which are nearly uniform in the longitudinal direction, may  be considered as
cylindrical perturbations in the energy density  upon a continuous background as depicted in Fig. 1.
The propagation of perturbations on the top of a  QGP background   has been investigated in several works
\cite{shuryak1,shuryak2,fn5,fn7}. In most of these works
\cite{shuryak1,shuryak2} a linearized version of hydrodynamics is employed. We have tried to keep the
nonlinear terms in the equations which describe the evolution of the perturbations \cite{fn5,fn7}. This extends the
validity of our formalism to perturbations which are not so small.

\begin{figure}[h]
\begin{center}
\epsfig{file=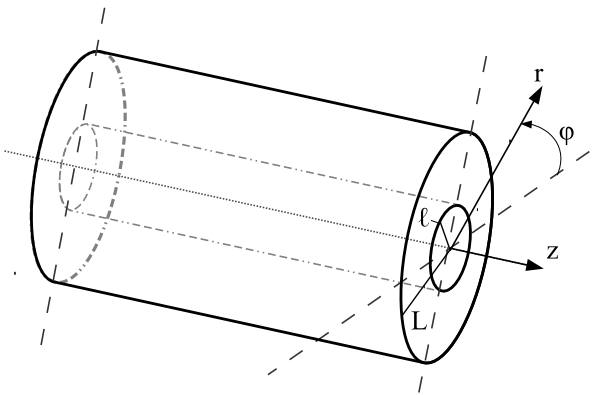,width=86mm}
\caption{Tubular perturbations on a QGP background.  The inner cylinder of radius $l$ represents a tube of energy
density higher than the background, shown as a cylindrical fireball of radius $L$. The perturbation expands
radially.  }
\end{center}
\label{fig1}
\end{figure}

In this work we try to answer the question: how fast do the tubes expand in the QGP ? In order to obtain the
answer we write the hydrodynamical equations for the propagation of cylindrical perturbations along the
radial direction (see Fig. 1), solve them numerically, and estimate what is the time needed for a tube
of initial radius of the order of  1 fm to grow and reach the typical radius of the system formed in heavy ion collisions,
which is of the order of 7 fm. If the tube expansion time were much shorter than the lifetime of the fireball, then the tube would
be very rapidly incorporated in the fireball and it would produce no visible effect in the final state particle correlation measurements.

We also investigate the effects of viscosity on the expansion of the tubes. It is well known that the relativistic version of the Navier-Stokes
equation does not constitute a causal theory.  We strongly recommend the reading of \cite{roma} to understand the subject in details.
Besides these conceptual issues of stability and causality we
perform a Navier-Stokes approach without worrying about microscopic time scales, due to the nonlinear expansion as seen in the appendix.
A future and more complete version of this work in relativistic viscous hydrodynamics is in progress with the use of
M\"uller-Israel-Stewart theory, which is a causal theory  \cite{roma}.

Due to dissipation, viscosity damps the perturbations, which
are then more easily mixed with the background fluid, loosing their influence on final state particle correlations.

In contrast to other studies of perturbations in fluids, we do not neglect the
nonlinear terms in the hydrodynamical equations.

This text is organized as follows.
In the next section we review the basic formulas of relativistic hydrodynamics. In section III we review
the equation of state (EOS) of the MIT bag model. In section IV we derive the equation which describes the evolution
of the tube. In section V we solve this equation numerically and  present some conclusions.

\section{Relativistic Fluid Dynamics}

Pedagogical texts on relativistic hydrodynamics can be found in \cite{wein,land}. Approximation
schemes which conserve nonlinearities can be found in \cite{davidson} and their application to
the study of  nonlinear waves in cold and warm nuclear matter can be found in \cite{fn1,fn2,fn3,fn4}
and references therein. In this section  we briefly review the basic equations (throughout this work
we use $c=1$, $\hbar=1$ and the Boltzmann constant is taken to be one, i.e., $k_{B}=1$).

For simplicity we start our discussion considering two coaxial cylinders. The inner and narrower cylinder
represents the  flux tube which is a perturbation in energy density $\varepsilon$. The outer and larger cylinder
represents the fireball with a uniform energy  density $\varepsilon_0$ ($ \varepsilon_0  \leq  \varepsilon$).
We will study the expansion of the flux tube in the center of mass system of the fireball. It is then natural to
chose spatial cylindrical coordinates $(z,r,\phi)$.

The  velocity  four-vector $u^{\nu}$ is defined as $u^{0}=\gamma$,  $\vec{u}=\gamma \vec{v}$,
where $\gamma$ is the Lorentz factor $\gamma=(1-v^{2})^{-1/2}$ and thus $u^{\nu}u_{\nu}=1$.
The velocity field of matter is given by  $\vec{v}=\vec{v}(t,r,z,\phi)$.  Because of the azimuthal symmetry
we do not have components along the $\phi$ direction and consequently  no terms involving $\partial/ \partial \phi$
will survive in what follows.

\subsection{Ideal fluid}

The energy momentum tensor is given by:
\begin{equation}
T_{\mu \nu}=(\varepsilon +p)u_{\mu}u_{\nu}-pg_{\mu\nu}
\label{tensor}
\end{equation}
where  $\varepsilon$ is the energy density and $p$ the pressure.
Energy-momentum conservation is given by:
\begin{equation}
\partial_{\nu}{T_{\mu}}^{\nu}=0
\label{cons}
\end{equation}
The projection of  (\ref{cons}) on a direction perpendicular to $u^{\mu}$ yields the relativistic version of
 Euler equation:
\begin{equation}
{\frac{\partial {\vec{v}}}{\partial t}}+(\vec{v} \cdot \vec{\nabla})\vec{v}=
-{\frac{1}{(\varepsilon + p)\gamma^{2}}}
\bigg({\vec{\nabla} p +\vec{v} {\frac{\partial p}{\partial t}}}\bigg)
\label{eul}
\end{equation}
The relativistic version of the continuity equation for the entropy density comes from the  projection of
(\ref{cons}) on the direction of  $u^{\nu}$:
\begin{equation}
(\varepsilon+p)\partial_{\mu}u^{\mu}+u^{\mu}\partial_{\mu}\varepsilon=0
\label{gconspro}
\end{equation}
We next recall  the  Gibbs relation:
\begin{equation}
\varepsilon + p={\mu_{B}}{\rho_{B}}+Ts
\label{gibaoporV}
\end{equation}
and the first law of thermodynamics:
\begin{equation}
d\varepsilon=Tds+{\mu_{B}}d{\rho_{B}}
\label{primlei}
\end{equation}
In the central rapidity region of  heavy ion collisions we expect to find hot QGP with zero net baryon number and
hence $\rho_{B}=0$  and  $T\neq 0$. Using   $d\rho_{B}=0$ in (\ref{primlei}) and inserting  (\ref{primlei}) and
(\ref{gibaoporV}) into (\ref{gconspro}) we find:
$$
Ts(\partial_{\mu}u^{\mu})+Tu^{\mu}(\partial_{\mu}s)=0
$$
and finally:

\begin{equation}
\partial_{\nu}(s{u}^{\nu})=0
\label{scons}
\end{equation}
as expected for a perfect fluid. This expression can be rewritten as:
\begin{equation}
{\frac{\partial s}{\partial t}}+\gamma^{2}v s\Bigg({\frac{\partial v}
{\partial t}}+ \vec{v}\cdot \vec{\nabla} v\Bigg)+\vec{\nabla} \cdot (s\vec{v})=0
\label{sconsc}
\end{equation}

\subsection{Viscous fluid}

In order to take the viscosity into account, we add the viscous stress tensor $\Pi^{\mu\nu}$  to the
ideal fluid energy-momentum tensor:
\begin{equation}
T^{\mu\nu} = T_{(0)}^{\mu\nu}+\Pi^{\mu\nu}
\label{emtensor}
\end{equation}
where $T_{(0)}$
is the ideal  fluid energy-momentum tensor \cite{wein,land}.  With this new definition of the energy-momentum tensor,
Eq. (\ref{cons}) remains valid.  We will  consider \cite{roma} a system without conserved charges (or at zero chemical
potential). As in the case of the  ideal fluid, we take the
appropriate projections of  (\ref{cons}), which are parallel
$(u_{\nu}\partial_{\mu}T^{\mu\nu})$ and perpendicular $(\Delta^{\alpha}_{\nu}\partial_{\mu}T^{\mu\nu})$
 to the fluid velocity obtaining:
\begin{equation}
u_{\nu}\partial_{\mu}T^{\mu\nu}=D\varepsilon + (\varepsilon+p)\partial_{\mu}u^{\mu}
+u_{\nu}\partial_{\mu}\Pi^{\mu\nu}=0
\label{afv1}
\end{equation}
and
\begin{equation}
\Delta^{\alpha}_{\nu}\partial_{\mu}T^{\mu\nu}=(\varepsilon+p)Du^{\alpha} - \nabla^{\alpha}p
+\Delta^{\alpha}_{\nu}\partial_{\mu}\Pi^{\mu\nu}=0
\label{afv2}
\end{equation}
where  $D \equiv u^{\mu}\partial_{\mu}$ \, and \, $\Delta^{\mu\nu} = g^{\mu\nu}-u^{\mu}u^{\nu}$.
The  viscous tensor is given by \cite{roma}:
\begin{equation}
\Pi^{\mu\nu}=\eta \nabla^{\langle\mu}u^{\nu\rangle} +\zeta \Delta^{\mu\nu} \nabla_{\alpha}u^{\alpha}
\label{vistsplittensor}
\end{equation}
where \cite{roma}
\begin{equation}
\nabla_{\langle\mu}u_{\nu\rangle} \equiv 2\nabla_{(\mu}u_{\nu)}-{\frac{2}{3}}\Delta_{\mu\nu}\nabla_{\alpha}u^{\alpha}
\label{tracelessnot}
\end{equation}
with
\begin{equation}
A_{(\mu}B_{\nu)}={\frac{1}{2}}(A_{\mu}B_{\nu}+A_{\nu}B_{\mu})
\label{symm}
\end{equation}
and
\begin{equation}
\nabla^{\alpha} \equiv \Delta^{\mu\alpha} \partial_{\mu}
\label{opers}
\end{equation}
Combining Eqs. (\ref{afv1}) and (\ref{afv2}) we can obtain the relativistic version of the Navier-Stokes equation. In a compact
form it may be found in \cite{roma}. For our purposes it is more convenient to write it in  the long form:
$$
(\varepsilon+p)\gamma^{2} \bigg({\frac{\partial}{\partial t}}+\vec{v} \cdot \vec{\nabla} \bigg) \vec{v}+
\vec{v}{\frac{\partial p}{\partial t}} + \vec{\nabla}p
$$
$$
-\eta \vec{v} \bigg\{\partial_{\mu}\partial^{\mu}\gamma+\partial_{\mu}{\frac{\partial u^{\mu}}{\partial t}}-\partial_{\mu}\bigg[\gamma\bigg({\frac{\partial}{\partial t}}+\vec{v} \cdot \vec{\nabla}\bigg)(\gamma u^{\mu}) \bigg]\bigg\}
-\vec{v}\bigg(\zeta-{\frac{2}{3}}\eta \bigg){\frac{\partial}{\partial t}}\bigg[
{\frac{\partial \gamma}{\partial t}}+\vec{\nabla}\cdot (\gamma \vec{v}) \bigg]
$$
$$
+\vec{v}\bigg(\zeta-{\frac{2}{3}}\eta \bigg)\partial_{\mu} \bigg\{ \gamma u^{\mu} \bigg[{\frac{\partial \gamma}{\partial t}}+\vec{\nabla}\cdot (\gamma \vec{\
v}) \bigg]\bigg\}
$$
$$
+\eta \bigg\{\partial_{\mu}\partial^{\mu}(\gamma \vec{v})-\partial_{\mu}\vec{\nabla}u^{\mu}
-\partial_{\mu} \bigg[ \gamma\bigg({\frac{\partial}{\partial t}}+\vec{v} \cdot \vec{\nabla}\bigg)(\gamma \vec{v} u^{\mu}) \bigg] \bigg\}
-\bigg( \zeta-{\frac{2}{3}}\eta \bigg)\vec{\nabla} \bigg[ {\frac{\partial \gamma}{\partial t}}+\vec{\nabla}\cdot (\gamma \vec{v}) \bigg]
$$
\begin{equation}
-\bigg(\zeta-{\frac{2}{3}}\eta \bigg)\partial_{\mu} \bigg\{ \gamma \vec{v} u^{\mu} \bigg[{\frac{\partial \gamma}{\partial t}}+\vec{\nabla}\cdot (\gamma \vec{v}) \bigg] \bigg\}=0
\label{rnsagain}
\end{equation}
With the help of  Eqs. (\ref{afv1}) and (\ref{afv2}) and using  thermodynamical relations we obtain \cite{roma}:
\begin{equation}
\partial_{\mu} s^{\mu}={\frac{1}{T}}\Pi^{\mu\nu}\nabla_{(\mu}u_{\nu)}
\label{seclawr}
\end{equation}
For our purposes we shall rewrite it as:
$$
\gamma {\frac{\partial s}{\partial t}}+ \gamma \vec{\nabla}s \cdot \vec{v} + s{\frac{\partial \gamma}{\partial t}}+
s\vec{\nabla}\gamma \cdot \vec{v}+\gamma s \vec{\nabla} \cdot \vec{v}=
-{\frac{\eta}{T}}\bigg({\frac{\partial \gamma}{\partial t}}\bigg)^{2}
-2{\frac{\eta}{T}}\bigg[\vec{\nabla} \gamma \cdot {\frac{\partial}{\partial t}}(\gamma\vec{v})
\bigg]
$$
\begin{equation}
-{\frac{\eta}{T}}(\partial^{i}u^{j})\partial_{j}u_{i}
+{\frac{1}{T}}\bigg({\frac{2}{3}}\eta+\zeta \bigg) \, \bigg[ {\frac{\partial \gamma}{\partial t}}
+\gamma  \vec{\nabla} \cdot \vec{v} +\vec{\nabla}\gamma \cdot \vec{v} \bigg]^{2}
\label{relcontss}
\end{equation}
which is the relativistic version of the continuity equation for the entropy density $s$.  In the case of an ideal fluid
($\eta=\zeta=0$) we recover the entropy density conservation:
$$
\gamma {\frac{\partial s}{\partial t}}+ \gamma \vec{\nabla}s \cdot \vec{v} +
s{\frac{\partial \gamma}{\partial t}}+s\vec{\nabla}\gamma \cdot \vec{v}+\gamma s \vec{\nabla} \cdot \vec{v}=0
$$

\section{Equation of state}

From the thermodynamics of the MIT bag model we have \cite{fn5}:

\begin{equation}
3(p+\mathcal{B})=\varepsilon-\mathcal{B}={\frac{8\pi^{2}}{15}}
\ T^{4}+{\frac{6}{\pi^{2}}}\int_{0}^{\infty} d{k}\hspace{0.2cm}k^{3}
[n_{\vec{k}}+\bar{n}_{\vec{k}}]
\label{bacana}
\end{equation}
and
\begin{equation}
p={\frac{1}{3}}\varepsilon-{\frac{4}{3}}\mathcal{B}
\label{eosqg}
\end{equation}
with the speed of sound  $c_{s}$, given by:
\begin{equation}
{c_{s}}^{2}={\frac{\partial p}{\partial \varepsilon}}={\frac{1}{3}}
\label{soundone}
\end{equation}
Since  $\rho_{B}=0$, the chemical potential  is zero$(\mu=0)$  and the distribution functions  are the same
for quarks and anti-quarks: ${n}_{\vec{k}}=\bar{n}_{\vec{k}} = {1}/{(1+e^{k/ T})}$. Therefore:
\begin{equation}
3(p+\mathcal{B})=\varepsilon-\mathcal{B}={\frac{37}{30}} \pi^{2} T^{4}
\label{bacanaaend}
\end{equation}
Solving the first identity for the pressure and using  the relation $s= ({{\partial p}/{\partial T}})_{V}$
we arrive at:
\begin{equation}
s={\frac{\partial }{\partial T}}\bigg(-\mathcal{B}+{\frac{37}{90}} \pi^{2} T^{4}\bigg)=
4  \\ {\frac{37}{90}} \pi^{2} T^{3}
\label{denstemp}
\end{equation}
The bag constant is related to the critical temperature, $T_c$, of the quark-hadron phase transition.
During the phase transition the pressure remains constant and (\ref{bacanaaend}) reduces to:
$$
\mathcal{B}={\frac{37}{30}} \pi^{2} {\frac{{T_{c}}^{4}}{3}} - const=
{\frac{37}{30}} \pi^{2}\bigg[{\frac{{T_{c}}^{4}}{3}}-{\frac{30}{37 \pi^{2}}}const\bigg]
$$
and we can define
$$
{T_{B}}^{4}=\bigg[{\frac{{T_{c}}^{4}}{3}}-{\frac{30}{37 \pi^{2}}}const\bigg]
$$
and consequently:
\begin{equation}
\mathcal{B}={\frac{37}{30}} \pi^{2} (T_{B})^{4}
\label{BT}
\end{equation}
The bag constant, $\mathcal{B}$,  is chosen to be $\mathcal{B}^{1/4}=170MeV$ and this corresponds to
$T_B=91 \, MeV$.
Inserting  (\ref{BT})  into the second identity of (\ref{bacanaaend}) we find the following expression for
$\varepsilon(T)$:
\begin{equation}
\varepsilon={\frac{37}{30}} \pi^{2}\Bigg( T^{4} + {T_{B}}^{4} \Bigg)
\label{deneb}
\end{equation}
Solving the second identity in (\ref{bacanaaend}) for the temperature, we find:
\begin{equation}
T=\Bigg[{\frac{30}{37\pi^{2}}}(\varepsilon-\mathcal{B})\Bigg]^{1/4}
\label{TfromEpsilon}
\end{equation}
which substituted in (\ref{denstemp}) yields:
\begin{equation}
s=s(\varepsilon)=4  \\ {\frac{37}{90}} \pi^{2} \Bigg[{\frac{30}{37\pi^{2}}}
(\varepsilon-\mathcal{B})\Bigg]^{3/4}
\label{densenerd}
\end{equation}
From (\ref{bacanaaend}) we have:
\begin{equation}
\varepsilon+p={\frac{148}{90}} \pi^{2} T^{4}
\label{eps+pe}
\end{equation}
and from (\ref{soundone}):
\begin{equation}
\vec{\nabla}p={\frac{1}{3}}\vec{\nabla}\varepsilon \hspace{2cm} and  \hspace{2cm}
{\frac{\partial p}{\partial t}} = {\frac{1}{3}}{\frac{\partial \varepsilon}{\partial t}}
\label{grads}
\end{equation}

\section{The wave equation}

In this section we combine the results of the two previous sections and derive the differential equations which
govern the evolution of  cylindrical perturbations in the energy density. We start writing the energy density
and the components of the fluid velocity in a dimensionless form:
\begin{equation}
\hat{\varepsilon}={\frac{\varepsilon}{\varepsilon_{0}}}
\label{vadimaft}
\end{equation}
\begin{equation}
\hat v={\frac{v}{c_{s}}}
\label{vadimaftagge}
\end{equation}
and
\begin{equation}
\hat v_{r}={\frac{v_{r}}{c_{s}}}  \hspace{0.2cm}, \hspace{0.5cm} \hat v_{z}={\frac{v_{z}}{c_{s}}}
\label{vadimaftag}
\end{equation}

\subsection{Ideal fluid}

After the use of  relations (\ref{eps+pe}) to (\ref{vadimaftag})
the components of the Euler equation (\ref{eul}) along the $r$ and $z$ directions become:

\begin{equation}
c_{s}{\frac{\partial \hat{v_{r}}}{\partial t}}+{c_{s}}^{2}\hat{v_{r}}{\frac{\partial \hat{v_{r}}}{\partial r}}
+{c_{s}}^{2}\hat{v_{z}}{\frac{\partial \hat{v_{r}}}{\partial z}}=
{\frac{15({c_{s}}^{2}{\hat{v}}^{2}-1)\varepsilon_{0}}{74\pi^{2}T^{4}}}
\bigg({\frac{\partial \hat{\varepsilon}}{\partial r}}+
{c_{s}}\hat{v_{r}}{\frac{\partial \hat{\varepsilon}}{\partial t}}\bigg)
\label{eulr}
\end{equation}
and
\begin{equation}
c_{s}{\frac{\partial \hat{v_{z}}}{\partial t}}+{c_{s}}^{2}\hat{v_{r}}{\frac{\partial \hat{v_{z}}}{\partial r}}
+{c_{s}}^{2}\hat{v_{z}}{\frac{\partial \hat{v_{z}}}{\partial z}}=
{\frac{15({c_{s}}^{2}{\hat{v}}^{2}-1)\varepsilon_{0}}{74\pi^{2}T^{4}}}
\bigg({\frac{\partial \hat{\varepsilon}}{\partial z}}+
{c_{s}}\hat{v_{z}}{\frac{\partial \hat{\varepsilon}}{\partial t}}\bigg)
\label{eulz}
\end{equation}
Now, using (\ref{bacanaaend}), (\ref{densenerd}) and (\ref{vadimaft})
to  (\ref{vadimaftag}) we rewrite the continuity equation (\ref{sconsc}) as:
$$
(1-{c_{s}}^{2}{\hat{v}}^{2})\Bigg\lbrace\Bigg({\frac{45\varepsilon_{0}}{74\pi^{2} T^{4}}}\Bigg)
\Bigg[{\frac{\partial \hat{\varepsilon}}{\partial t}}+
{c_{s}}\hat{v_{r}}{\frac{\partial \hat{\varepsilon}}{\partial r}}
+{c_{s}}\hat{v_{z}}{\frac{\partial \hat{\varepsilon}}{\partial z}}\Bigg]+{\frac{{c_{s}}\hat{v_{r}}}{r}}+
{c_{s}}{\frac{\partial \hat{v_{r}}}{\partial r}}
+{c_{s}}{\frac{\partial \hat{v_{z}}}{\partial z}}\Bigg\rbrace
$$
$$
+{c_{s}}^{2}\hat{v_{r}}{\frac{\partial \hat{v_{r}}}{\partial t}}+
{c_{s}}^{2}\hat{v_{z}}{\frac{\partial \hat{v_{z}}}{\partial t}}+
{c_{s}}^{3}{\hat{v_{r}}}^{2}{\frac{\partial \hat{v_{r}}}{\partial r}}
$$
\begin{equation}
+{c_{s}}^{3}{\hat{v_{r}}}{\hat{v_{z}}}{\frac{\partial \hat{v_{z}}}{\partial r}}+
{c_{s}}^{3}{\hat{v_{z}}}{\hat{v_{r}}}{\frac{\partial \hat{v_{r}}}{\partial z}}+
{c_{s}}^{3}{\hat{v_{z}}}^{2}{\frac{\partial \hat{v_{z}}}{\partial z}}=0
\label{sconscumdimagain}
\end{equation}
Now we  combine (\ref{eulr}), (\ref{eulz}) and  (\ref{sconscumdimagain}) to find the wave equation. To this end we perform a
change of variables in  (\ref{eulr}), (\ref{eulz}) and (\ref{sconscumdimagain}), going from the $(r,z,t)$ space to the  $(R,Z,T)$
space by the reductive perturbation method \cite{kp2010,rpt1,rpt2}, through the introduction of  the ``stretched'' coordinates
\cite{kp2010}:
\begin{equation}
R={\frac{\sigma^{1/2}}{L}}(r-{c_{s}}t)
\hspace{0.2cm}, \hspace{0.5cm}
Z={\frac{\sigma}{L}}z
\hspace{0.2cm}, \hspace{0.5cm}
T={\frac{{\sigma^{3/2}}}{L}}{c_{s}}t
\label{streta}
\end{equation}
where $L$ is a characteristic length scale of the problem (typically the radius of a heavy ion) and $\sigma$ is a small
expansion parameter.  We next perform the following expansions \cite{kp2010,rpt1,rpt2}:
\begin{equation}
\hat\varepsilon=1+\sigma \varepsilon_{1}+ \sigma^{2} \varepsilon_{2}
+\sigma^{3} \varepsilon_{3}+ \dots
\label{roexpa}
\end{equation}
\begin{equation}
\hat {v_{r}}=\sigma v_{{r}_1}+ \sigma^{2} v_{{r}_2}+ \sigma^{3} v_{{r}_3}+ \dots
\label{vexpa}
\end{equation}
\begin{equation}
\hat {v_{z}}=\sigma^{3/2} v_{{z}_1}+ \sigma^{5/2} v_{{z}_2}+ \sigma^{7/2} v_{{z}_3}+ \dots
\label{vexpar}
\end{equation}
After the use of (\ref{streta}), (\ref{roexpa}), (\ref{vexpa}) and (\ref{vexpar}),
the Euler and continuity equations
can be written as  series in powers of $\sigma$. We will consider terms only up to the order $\sigma^{2}$.
It is then possible to reorganize the series in powers of $\sigma$, $\sigma^{3/2}$ and  $\sigma^{2}$.
After  a little algebra we find:
$$
\sigma \Bigg\{ \Bigg({\frac{45\varepsilon_{0}}{74\pi^{2}T^{4}}}\Bigg){\frac{\partial \varepsilon_{1}}
{\partial R}} -{\frac{\partial v_{{r}_1}}
{\partial R}}\Bigg\}
$$
\begin{equation}
+\sigma^{2}\Bigg\{\Bigg({\frac{45\varepsilon_{0}}{74\pi^{2}T^{4}}}\Bigg){\frac{\partial \varepsilon_{2}}
{\partial R}} -{\frac{\partial v_{{r}_2}}
{\partial R}}+{\frac{\partial v_{{r}_1}}{\partial T}}+v_{{r}_1}{\frac{\partial v_{{r}_1}}{\partial R}}
-\Bigg({\frac{15\varepsilon_{0}}{74\pi^{2}T^{4}}}\Bigg)v_{{r}_1}{\frac{\partial \varepsilon_{1}}{\partial R}}\Bigg\}=0
\label{sigma1}
\end{equation}
\begin{equation}
\sigma^{3/2}\Bigg\{\Bigg({\frac{45\varepsilon_{0}}{74\pi^{2}T^{4}}}\Bigg){\frac{\partial \varepsilon_{1}}
{\partial Z}} -{\frac{\partial v_{{z}_1}}
{\partial R}}\Bigg\}=0
\label{sigma2}
\end{equation}
and
$$
\sigma \Bigg\{ -\Bigg({\frac{45\varepsilon_{0}}{74\pi^{2}T^{4}}}\Bigg){\frac{\partial \varepsilon_{1}}
{\partial R}} +{\frac{\partial v_{{r}_1}}
{\partial R}}\Bigg\}
$$
$$
+\sigma^{2}\Bigg\{-\Bigg({\frac{45\varepsilon_{0}}{74\pi^{2}T^{4}}}\Bigg){\frac{\partial \varepsilon_{2}}
{\partial R}}+\Bigg({\frac{45\varepsilon_{0}}{74\pi^{2}T^{4}}}\Bigg){\frac{\partial \varepsilon_{1}}
{\partial T}} +\Bigg({\frac{45\varepsilon_{0}}{74\pi^{2}T^{4}}}\Bigg)v_{{r}_1}{\frac{\partial \varepsilon_{1}}{\partial R\
}}
$$
\begin{equation}
+{\frac{v_{{r}_1}}{T}}+{\frac{\partial v_{{r}_2}}{\partial R}}
+{\frac{\partial v_{{z}_1}}{\partial Z}}-{\frac{v_{{r}_1}}{3}}{\frac{\partial v_{{r}_1}}{\partial R}}\Bigg\}=0
\label{sigma3}
\end{equation}
In the above equations each bracket must vanish independently and therefore we obtain a set of relations. From the terms of
order $\sigma$ in the last two equations we find:
\begin{equation}
\Bigg({\frac{45\varepsilon_{0}}{74\pi^{2}T^{4}}}\Bigg){\frac{\partial \varepsilon_{1}}
{\partial R}}={\frac{\partial v_{{r}_1}}{\partial R}}
\label{sigmar1}
\end{equation}
which, after the integration over $R$ and taking the integration constant equal to zero, yields:
\begin{equation}
v_{{r}_1}=\Bigg({\frac{45\varepsilon_{0}}{74\pi^{2}T^{4}}}\Bigg) \varepsilon_{1}
\label{sigmar1a}
\end{equation}
From the terms of order  $\sigma^{3/2}$ we have:
\begin{equation}
\Bigg({\frac{45\varepsilon_{0}}{74\pi^{2}T^{4}}}\Bigg){\frac{\partial \varepsilon_{1}}
{\partial Z}}={\frac{\partial v_{{z}_1}}
{\partial R}}
\label{sigmar2}
\end{equation}
which, after the derivation with respect to $Z$, becomes:
\begin{equation}
\Bigg({\frac{45\varepsilon_{0}}{74\pi^{2}T^{4}}}\Bigg){\frac{\partial^{2} \varepsilon_{1}}
{\partial Z^{2}}}={\frac{\partial^{2} v_{{z}_1}}{\partial Z \partial R}}
\label{sigmar2a}
\end{equation}
From the terms of order $\sigma^{2}$ we obtain:
\begin{equation}
{\frac{\partial v_{{r}_2}}{\partial R}}
-\Bigg({\frac{45\varepsilon_{0}}{74\pi^{2}T^{4}}}\Bigg){\frac{\partial \varepsilon_{2}}
{\partial R}}={\frac{\partial v_{{r}_1}}{\partial T}}+v_{{r}_1}{\frac{\partial v_{{r}_1}}{\partial R}} -
\Bigg({\frac{15\varepsilon_{0}}{74\pi^{2}T^{4}}}\Bigg)v_{{r}_1}{\frac{\partial
\varepsilon_{1}}{\partial R}}
\label{sigmar3}
\end{equation}
and
$$
{\frac{\partial v_{{r}_2}}{\partial R}}
-\Bigg({\frac{45\varepsilon_{0}}{74\pi^{2}T^{4}}}\Bigg){\frac{\partial \varepsilon_{2}}
{\partial R}}=-\Bigg({\frac{45\varepsilon_{0}}{74\pi^{2}T^{4}}}\Bigg){\frac{\partial \varepsilon_{1}}
{\partial T}}-\Bigg({\frac{45\varepsilon_{0}}{74\pi^{2}T^{4}}}\Bigg)v_{{r}_1}{\frac{\partial \varepsilon_{1}}{\partial R}}
$$
\begin{equation}
-{\frac{v_{{r}_1}}{T}}-{\frac{\partial v_{{z}_1}}{\partial Z}}+{\frac{v_{{r}_1}}{3}}{\frac{\partial v_{{r}_1}}{\partial R}}
\label{sigmar4}
\end{equation}
Identifying  (\ref{sigmar3}) with  (\ref{sigmar4}), using (\ref{sigmar1a}), derivating the resulting equation with respect to
$R$ and using (\ref{sigmar2a}), we obtain
\begin{equation}
{\frac{\partial}{\partial R}}\Bigg\{{\frac{\partial\varepsilon_{1}}{\partial T}}+
\Bigg[\frac{2}{3}\Bigg({\frac{45\varepsilon_{0}}{74\pi^{2}T^{4}}}\Bigg)\Bigg]
\varepsilon_{1}{\frac{\partial \varepsilon_{1}}{\partial R}}
+{\frac{\varepsilon_{1}}{2T}}\Bigg\}
+{\frac{1}{2}}{\frac{\partial^{2} \varepsilon_{1}}{\partial Z^{2}}}=0
\label{bwqcdxitauedp}
\end{equation}
Returning now to the  $(r,z,t)$ space we find:
\begin{equation}
{\frac{\partial}{\partial r}}\Bigg\{{\frac{\partial\hat\varepsilon_{1}}{\partial t}}+
c_{s}{\frac{\partial\hat\varepsilon_{1}}{\partial r}}+
\Bigg[\frac{2}{3}\Bigg({\frac{45\varepsilon_{0}}{74\pi^{2}T^{4}}}\Bigg)\Bigg]c_{s}
\ \hat\varepsilon_{1}{\frac{\partial \hat\varepsilon_{1}}{\partial r}}
+{\frac{\hat\varepsilon_{1}}{2t}}\Bigg\}
+{\frac{c_{s}}{2}}{\frac{\partial^{2} \hat\varepsilon_{1}}{\partial z^{2}}}=0
\label{bwqcdTfin}
\end{equation}
where $\hat\varepsilon_{1}\equiv \sigma\varepsilon_{1}$ is a small perturbation on the background energy density
$\varepsilon_{0}$.
We can rewrite the above equation with the coefficients depending only on temperatures and on the sound velocity.
Using (\ref{deneb}) and calling $\varepsilon_0=\varepsilon (T=T_0)$,  where $T_{0}$  is the temperature of the
background, we then find:
\begin{equation}
{\frac{2}{3}}\Bigg({\frac{45\varepsilon_{0}}{74\pi^{2}{T_{0}}^{4}}}\Bigg)=
{\frac{1}{2}}\Bigg[1+ \Bigg({\frac{T_{B}}{T_{0}}}\Bigg)^{4}\Bigg]
\label{relaT}
\end{equation}
where $T_0 > T_B$.  Substituting  (\ref{relaT}) into (\ref{bwqcdTfin}) we find the final form of the wave equation:
\begin{equation}
{\frac{\partial}{\partial r}}\Bigg\{{\frac{\partial\hat\varepsilon_{1}}{\partial t}}+
c_{s}{\frac{\partial\hat\varepsilon_{1}}{\partial r}}+
{\frac{c_{s}}{2}}\Bigg[1+ \Bigg({\frac{T_{B}}{T_{0}}}\Bigg)^{4}\Bigg]
\ \hat\varepsilon_{1}{\frac{\partial \hat\varepsilon_{1}}{\partial r}}
+{\frac{\hat\varepsilon_{1}}{2t}}\Bigg\}
+{\frac{c_{s}}{2}}{\frac{\partial^{2} \hat\varepsilon_{1}}{\partial z^{2}}}=0
\label{bwqcdTfinest}
\end{equation}

\subsection{Viscous fluid}

Using again the  relations (\ref{eps+pe}) to (\ref{vadimaftag}) in
(\ref{rnsagain}), performing the same change of variables (\ref{streta}), performing the same expansions
(\ref{roexpa}),  (\ref{vexpa}) and (\ref{vexpar}), organizing the several terms in powers of $\sigma$ and obtaining the
corresponding identities and returning to the $(r, z, t)$ we  arrive at the analogue of (\ref{bwqcdTfin}) for a viscous
fluid:
\begin{equation}
{\frac{\partial}{\partial r}}\Bigg\{{\frac{\partial\hat\varepsilon_{1}}{\partial t}}+
c_{s}{\frac{\partial\hat\varepsilon_{1}}{\partial r}}+
\Bigg[\frac{2}{3}\Bigg({\frac{45\varepsilon_{0}}{74\pi^{2}{T_{0}}^{4}}}\Bigg)\Bigg]c_{s}
\ \hat\varepsilon_{1}{\frac{\partial \hat\varepsilon_{1}}{\partial r}}
+{\frac{\hat\varepsilon_{1}}{2t}}
-\Bigg({\frac{45}{74\pi^{2}{T_{0}}^{4}}}\Bigg) \Bigg(  \zeta + \frac{4}{3} \eta  \Bigg) \frac{\partial^{2} \hat\varepsilon_{1}}{\partial r^{2}}
\Bigg\}
+{\frac{c_{s}}{2}}{\frac{\partial^{2} \hat\varepsilon_{1}}{\partial z^{2}}}=0
\label{bwqcdTfin_v}
\end{equation}
As expected, the above equation reduces to the corresponding equation for ideal fluids (\ref{bwqcdTfin}) in the limit
$\eta = \zeta = 0$. Since the derivation of the equation is very similar to the sequence of steps that led to  (\ref{bwqcdTfin}), we
omitted all the details. However the interested reader can find some more details in the Appendix.

With the help of (\ref{denstemp}) we can rewrite the viscosity term as a function of the dimensionless ratios $\eta/s$ and $\zeta/s$, which are well studied in the literature \cite{chau,raju12}.  So the wave equation (\ref{bwqcdTfin_v}) becomes, after using (\ref{relaT}) and (\ref{denstemp}):
\begin{equation}
{\frac{\partial}{\partial r}}\Bigg\{{\frac{\partial\hat\varepsilon_{1}}{\partial t}}+
c_{s}{\frac{\partial\hat\varepsilon_{1}}{\partial r}}+
{\frac{c_{s}}{2}}\Bigg[1+ \Bigg({\frac{T_{B}}{T_{0}}}\Bigg)^{4}\Bigg]
\ \hat\varepsilon_{1}{\frac{\partial \hat\varepsilon_{1}}{\partial r}}
+{\frac{\hat\varepsilon_{1}}{2t}}
-{\frac{1}{T_{0}}}\Bigg({\frac{\zeta}{s}}  + \frac{4}{3} {\frac{\eta}{s}}  \Bigg) \frac{\partial^{2} \hat\varepsilon_{1}}{\partial r^{2}}
\Bigg\}
+{\frac{c_{s}}{2}}{\frac{\partial^{2} \hat\varepsilon_{1}}{\partial z^{2}}}=0
\label{bwqcdTfin_va}
\end{equation}

\subsection{Effect of the background expansion}

So far we have considered the motion of a perturbation on a static background. In order to include the motion of the
underlying medium we would need to know the full solution of the three-dimensional hydrodynamical equations describing
the QGP expansion and consequently $\varepsilon_0(r,\phi,z,t)$. The appearance of a coordinate dependent quantity in the
denominator of (\ref{vadimaft}) would make our expansion of the Euler and continuity equations too complicated.  A simple
way to estimate the effect of the expansion is to represent the cooling of the background by the Bjorken formula
\cite{bjorken}:
\begin{equation}
{\frac{T(\tau)}{T(\tau_{0})}}=\Bigg({\frac{\tau_{0}}{\tau}}\Bigg)^{1/3}
\label{bjorkent}
\end{equation}
where the proper time is given by $\tau={\frac{t}{\gamma}}=t\sqrt{1-v^{2}}$.
We have only radial flow $v^{2}={v_{r}}^{2}=(r/t)^{2}$ and thus:
\begin{equation}
\tau=\sqrt{t^{2}-r^{2}}
\label{propert}
\end{equation}
The initial proper time is taken to be $\tau_{0}=1$ fm. With the inclusion of  Bjorken cooling  the term in
parenthesis in (\ref{bwqcdTfinest}) will become:
\begin{equation}
\frac{T_B}{T_{0}} \rightarrow   \frac{T_B}{T_{0}(\tau)} = \frac{T_B}{T_{0}(\tau_0)} \Bigg(\frac{\tau}{\tau_0}\Bigg)^{1/3}
\label{ourto}
\end{equation}
Inserting (\ref{ourto}) into wave equation (\ref{bwqcdTfin_va})
we have:
$$
{\frac{\partial}{\partial r}}\Bigg\{{\frac{\partial\hat\varepsilon_{1}}{\partial t}}+
c_{s}{\frac{\partial\hat\varepsilon_{1}}{\partial r}}+
{\frac{c_{s}}{2}} \Bigg[ 1+ \Bigg( \frac{T_B}{T_{0}(\tau_0)}  \, \bigg(\frac{\tau}{\tau_0}\bigg)^{1/3} \, \, \Bigg)^{4} \Bigg]
\hat\varepsilon_{1}{\frac{\partial \hat\varepsilon_{1}}{\partial r}}
+{\frac{\hat\varepsilon_{1}}{2t}}
$$
\begin{equation}
-\frac{1}{T_{0}(\tau_0)} \Bigg(\frac{\tau}{\tau_0}\Bigg)^{1/3}
\Bigg({\frac{\zeta}{s}}  + \frac{4}{3} {\frac{\eta}{s}}  \Bigg) \frac{\partial^{2} \hat\varepsilon_{1}}{\partial r^{2}}\Bigg\}
+{\frac{c_{s}}{2}}{\frac{\partial^{2} \hat\varepsilon_{1}}{\partial z^{2}}}=0
\label{weqbv}
\end{equation}
for a viscous fluid. Inserting (\ref{ourto}) into (\ref{bwqcdTfinest}) we have:
\begin{equation}
{\frac{\partial}{\partial r}}\Bigg\{{\frac{\partial\hat\varepsilon_{1}}{\partial t}}+
c_{s}{\frac{\partial\hat\varepsilon_{1}}{\partial r}}+
{\frac{c_{s}}{2}} \Bigg[ 1+ \Bigg( \frac{T_B}{T_{0}(\tau_0)}  \, \bigg(\frac{\tau}{\tau_0}\bigg)^{1/3} \, \, \Bigg)^{4} \Bigg]
\hat\varepsilon_{1}{\frac{\partial \hat\varepsilon_{1}}{\partial r}}
+{\frac{\hat\varepsilon_{1}}{2t}}\Bigg\}
+{\frac{c_{s}}{2}}{\frac{\partial^{2} \hat\varepsilon_{1}}{\partial z^{2}}}=0
\label{weqbvi}
\end{equation}
for an ideal fluid.

\section{Numerical results and discussion}

For simplicity, we assume that when they are formed and also throughout the expansion the tubes are uniform
along the longitudinal direction and therefore:
$$
{\frac{c_{s}}{2}}{\frac{\partial^{2} \hat\varepsilon_{1}}{\partial z^{2}}} = 0
$$
Integrating (\ref{bwqcdTfinest}) and (\ref{bwqcdTfin_va}) with respect to $r$ and setting the integration constant to zero we arrive at
the cylindrical breaking wave equation for the ideal fluid:
\begin{equation}
{\frac{\partial\hat\varepsilon_{1}}{\partial t}}+
c_{s}{\frac{\partial\hat\varepsilon_{1}}{\partial r}}+
{\frac{c_{s}}{2}}\Bigg[1+ \Bigg({\frac{T_{B}}{T_{0}}}\Bigg)^{4}\Bigg]
\ \hat\varepsilon_{1}{\frac{\partial \hat\varepsilon_{1}}{\partial r}}
+{\frac{\hat\varepsilon_{1}}{2t}} =0
\label{weq}
\end{equation}
and at the famous Burgers equation \cite{kp2010,rpt1,rpt2} for the viscous fluid:
\begin{equation}
{\frac{\partial\hat\varepsilon_{1}}{\partial t}}+
c_{s}{\frac{\partial\hat\varepsilon_{1}}{\partial r}}+
{\frac{c_{s}}{2}}\Bigg[1+ \Bigg({\frac{T_{B}}{T_{0}}}\Bigg)^{4}\Bigg]
\ \hat\varepsilon_{1}{\frac{\partial \hat\varepsilon_{1}}{\partial r}}
+{\frac{\hat\varepsilon_{1}}{2t}}
={\frac{1}{T_{0}}}\Bigg({\frac{\zeta}{s}}  + \frac{4}{3} {\frac{\eta}{s}}  \Bigg) \frac{\partial^{2} \hat\varepsilon_{1}}{\partial r^{2}}
\label{burgers_final}
\end{equation}
which in this case is a cylindrical Burgers equation \cite{shukla,sahucb}.
Both can be solved numerically for a given choice of $T_0$ and $T_B$. $c^2_s=1/3$ for this equation of state.
If both equations had only the first two terms, they would describe a traveling wave with velocity $c_s$.
The third term makes the equations nonlinear in $\hat\varepsilon_{1}$. Its effect is to increase the velocity of
the wave, which is given by the coefficient of the terms proportional  to
${\partial \hat\varepsilon_{1}} / {\partial r}$. The velocity becomes therefore proportional to $\hat\varepsilon_{1}$
and so the top of the wave travels faster than its bottom. Because of this, an initially  gaussian pulse turns into a
triangular pulse with a ``vertical wall'', as it will be seen in the figures. Finally, the nonlinear term induces
rapid oscillations around the region close to the wall. This is called dispersion.
The term $\hat\varepsilon_{1}/2t$ in both equations comes
from the use of cylindrical geometry. It causes the attenuation of the initial perturbation at increasing times.  Changes
in the equation of state  imply changes in the evolution of the tube. A harder EOS will have a bigger velocity of sound
and this will make the tube move faster.  Moreover the strength of the nonlinear term  is directly proportional to $T_B$,
and consequently (because of (\ref{BT}) ) to the bag constant, which, in its turn, contains information about the
nonperturbative components of the EOS.  Increasing the bag constant makes the tube move faster! Inversely, increasing the
temperature of the background, $T_0$, makes the pulse to propagate slower.  In spite of the qualitative richness of
(\ref{weq})  and  (\ref{burgers_final}), for realistic values of $\mathcal{B}$ and $T_0$,
the nonlinear term has a very limited range of numerical values. Moreover, as we can observe in (\ref{weq}) and  in
(\ref{burgers_final}), this term is never large. Thus we can conclude a
posteriori that the linearization, as performed in \cite{shuryak1,shuryak2}, may indeed be a good approximation.
In the case of  the Burgers equation (\ref{burgers_final}) the second order derivative term tames the breaking and
dispersion of the wave and at the same time, dissipation reduces its amplitude.

The initial condition is given  by a gaussian pulse in $ \hat\varepsilon_{1}$:
\begin{equation}
\hat\varepsilon_{1} = A \, e^{-r^2 / r^2_0}
\label{condinit}
\end{equation}
where the amplitude $A$ and the approximate width $r_0$ are parameters which depend on the dynamics of flux tube formation.
For simplicity we shall refer to $r_0$ as the initial "radius" of the tube.
If the tubes are perturbations we expect that $A < 1$. According to current estimates \cite{raju12} the transverse size of the
tubes is of the order of $1$ fm and thus in our calculations $0.1 \, fm \leq r_0 \leq 0.8 \, fm$.
We consider hot QGP at temperatures $T_{0}= 150 \, MeV$ and $T_{0}=500 \, MeV$ treated as an ideal fluid
($\eta/s = \zeta/s = 0$) and as a viscous fluid ($\eta/s =0.08$ and $\zeta/s = 0$)  \cite{chau,raju12}.

In the numerical analysis there are many cases to be considered. We present our results in eight figures (Figs. 2 - 9).
The first four refer to the static background and the  second group of four shows solutions for the same parameters  for
the case of an expanding background. All figures have four panels. The two upper panels refer to the low temperature
($T_0=150$ MeV) and the two lower panels to the high temperature   ($T_0=500$ MeV). The two panels on the left show results
with the ideal fluid and the two panels on the right results with the viscous fluid. From the figures we want to see how
sensitive the expansion of the tube is to changes in:  i) the initial amplitude, ii)  the temperature, iii) the tube radius,
iv) the strength of viscosity and v) the expansion of the background fluid.   In what follows we discuss the role played by
each one of these variables mentioning them by order of relevance.

\subsubsection{Viscosity}

\begin{figure}[ht!]
\begin{center}
\subfigure[ ]{\label{fig:first}
\includegraphics[width=0.48\textwidth]{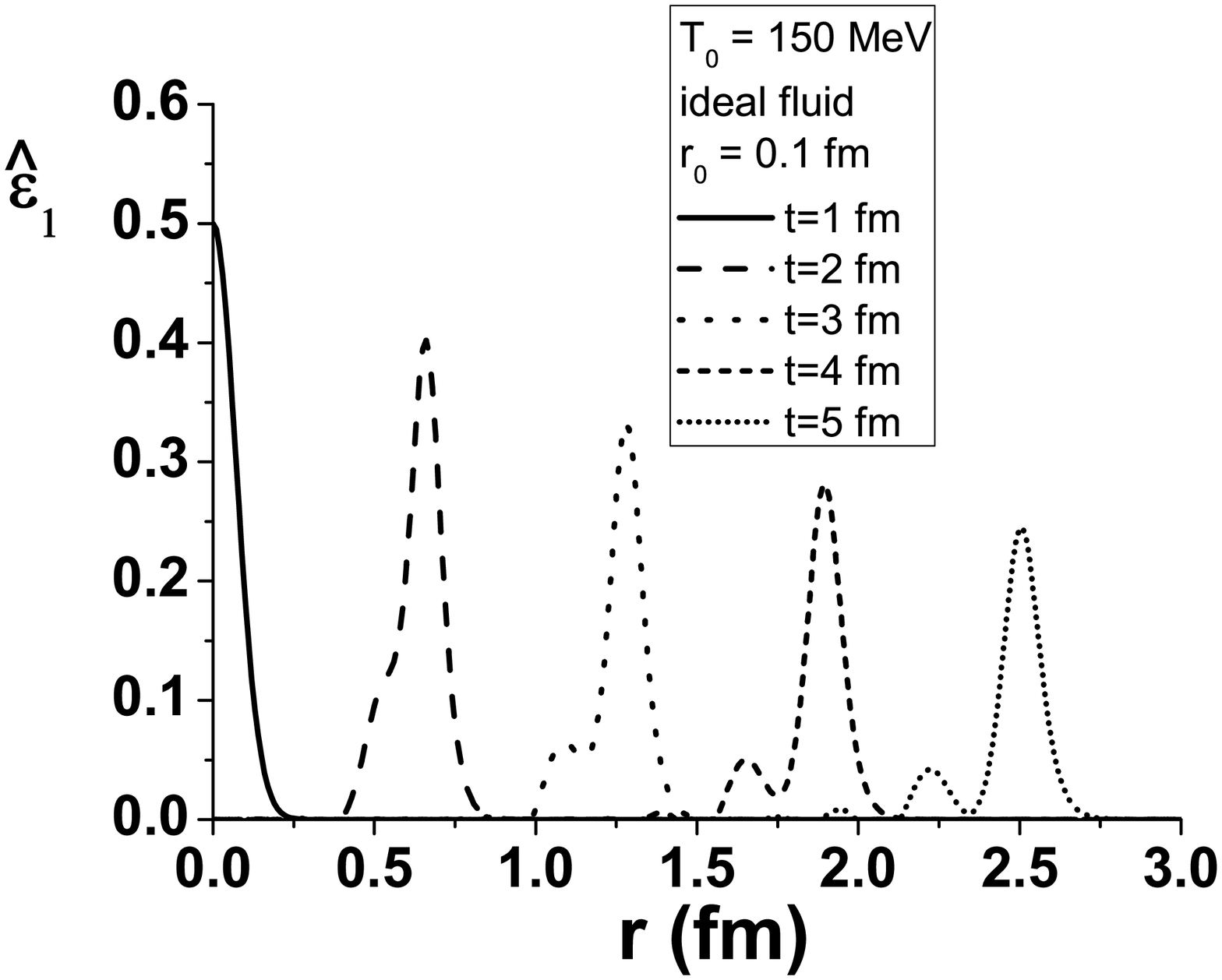}}
\subfigure[ ]{\label{fig:second}
\includegraphics[width=0.48\textwidth]{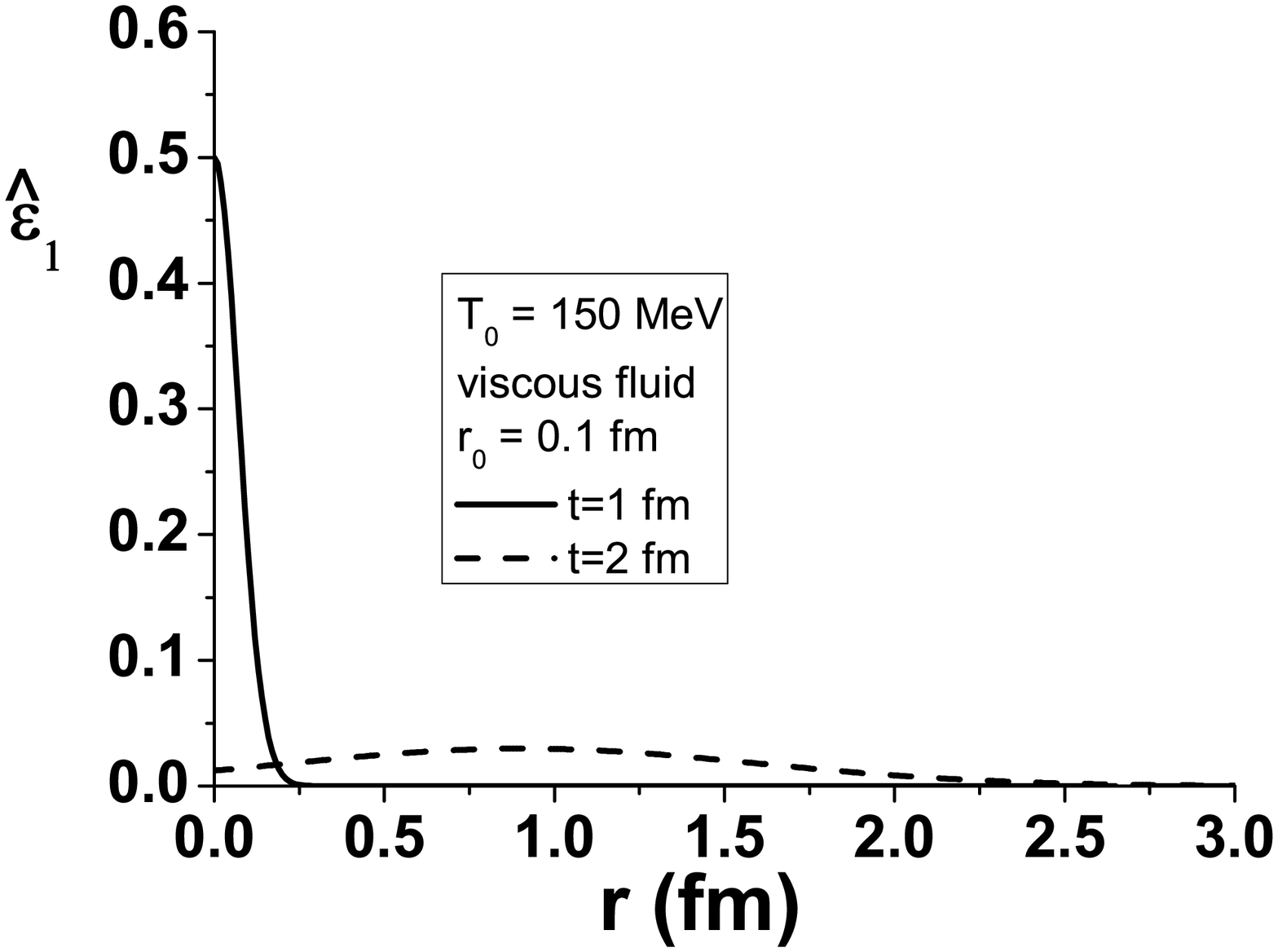}}\\
\subfigure[ ]{\label{fig:third}
\includegraphics[width=0.48\textwidth]{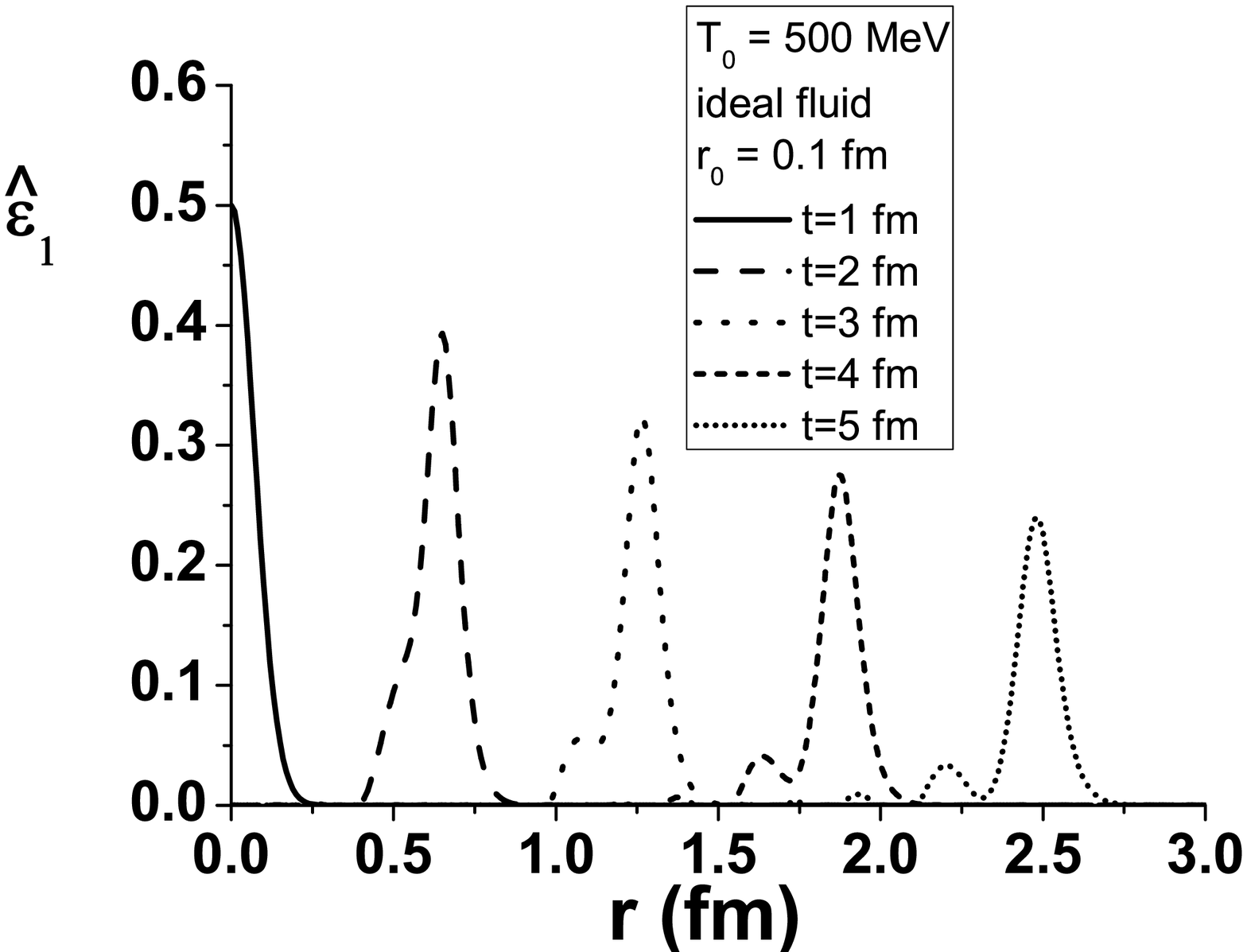}}
\subfigure[ ]{\label{fig:fourth}
\includegraphics[width=0.48\textwidth]{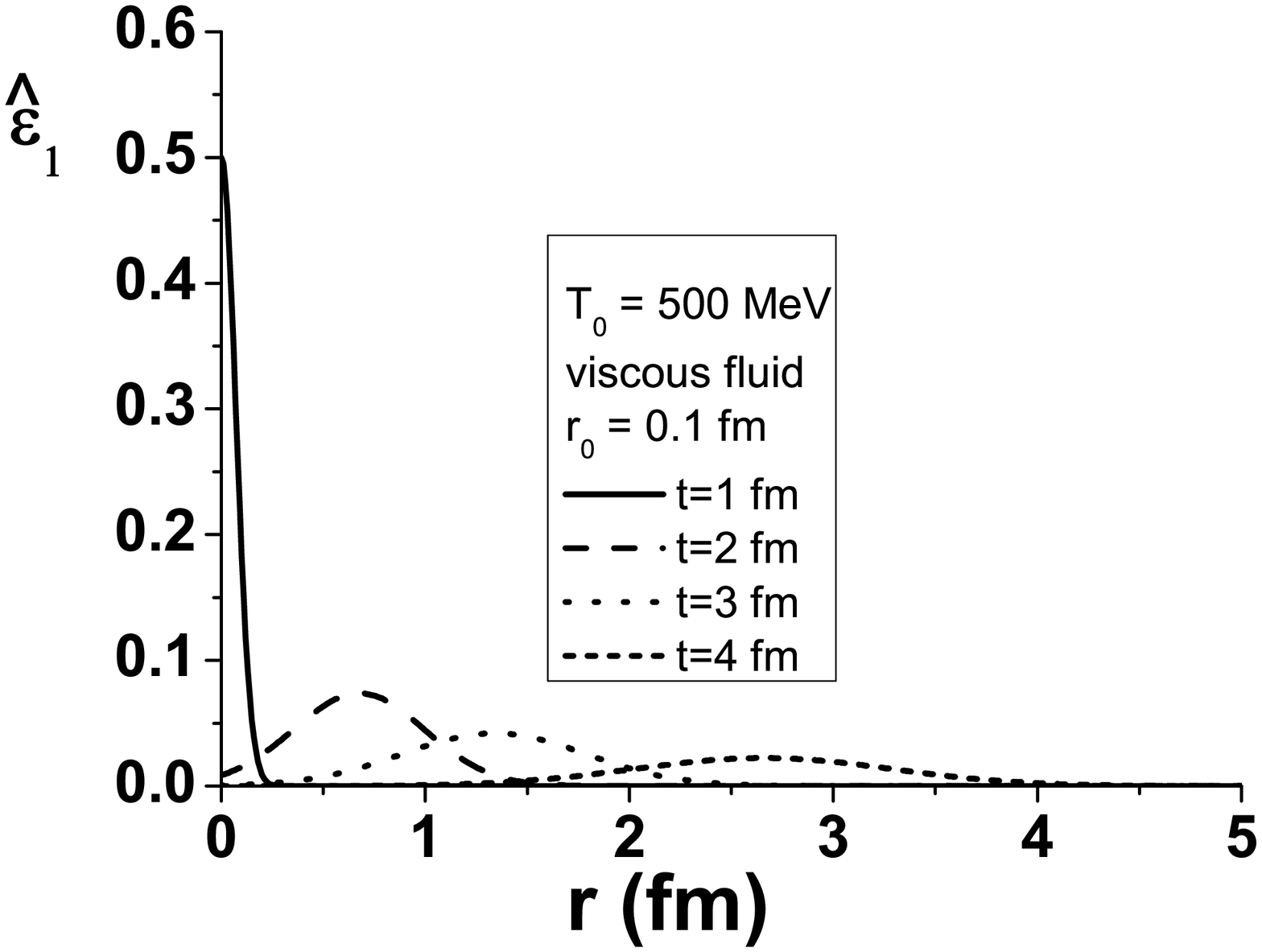}}
\end{center}
\caption{Numerical solutions of (\ref{weq}) for an  ideal fluid (left) and of
(\ref{burgers_final}) for a viscous fluid (right). The solutions are for several times and for $T_0=150$ MeV (upper panels)
and  $T_0=500$ MeV (lower panels). The initial  amplitude is $0.5$ and the width $r_{0}=0.1 \,\, fm$.}
\label{fig2}
\end{figure}

\begin{figure}[ht!]
\begin{center}
\subfigure[ ]{\label{fig:first}
\includegraphics[width=0.48\textwidth]{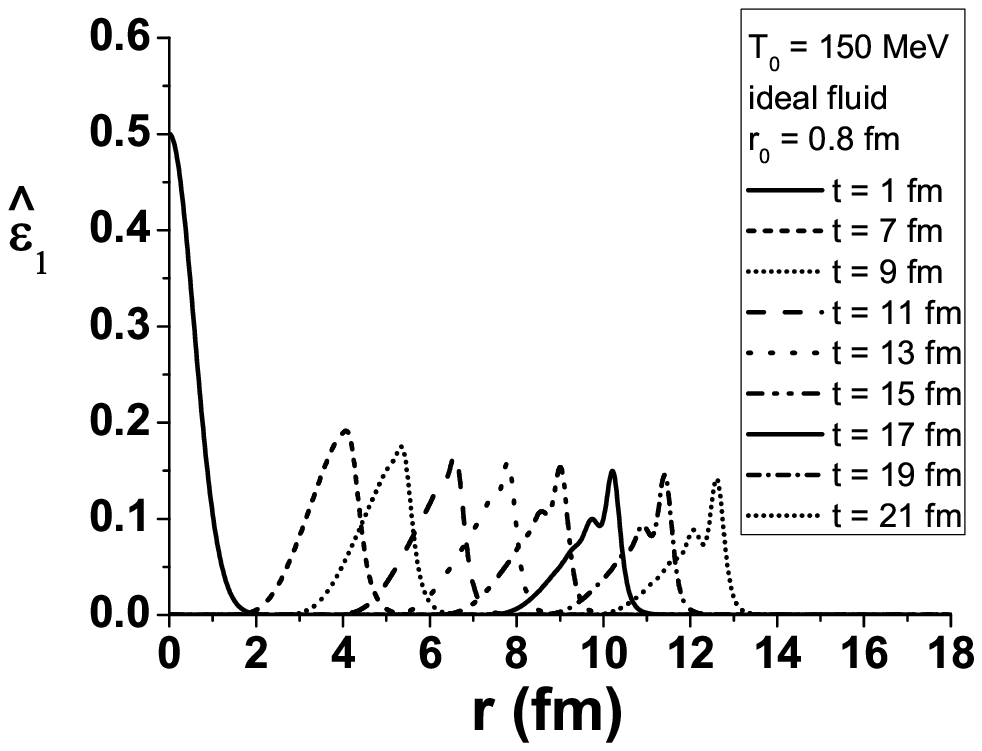}}
\subfigure[ ]{\label{fig:second}
\includegraphics[width=0.48\textwidth]{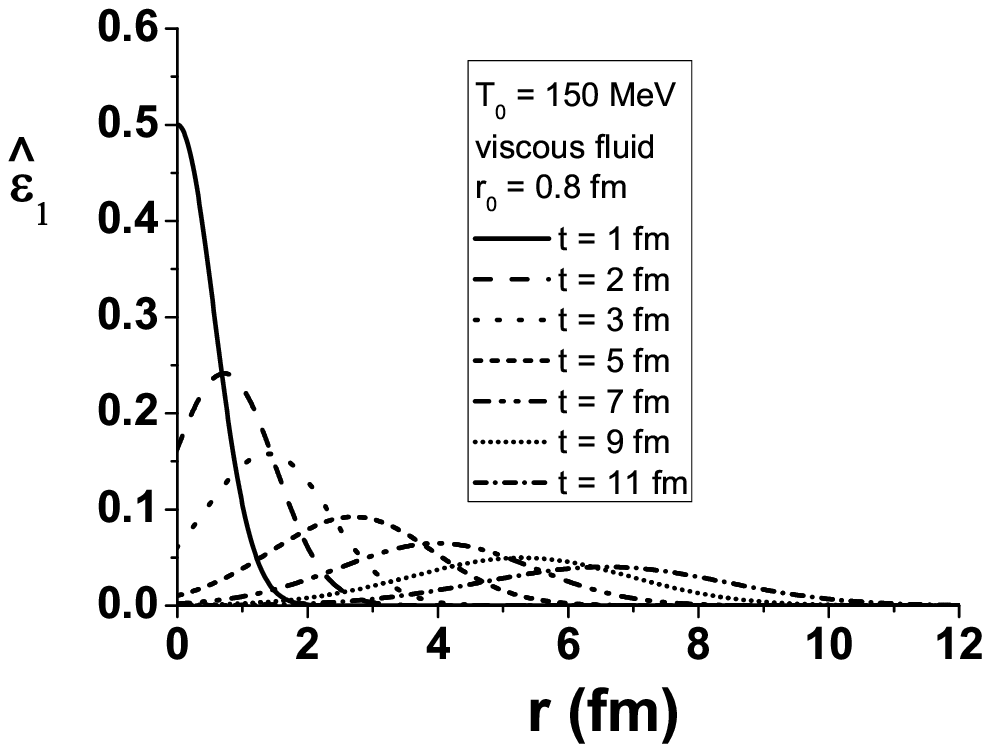}}\\
\subfigure[ ]{\label{fig:third}
\includegraphics[width=0.48\textwidth]{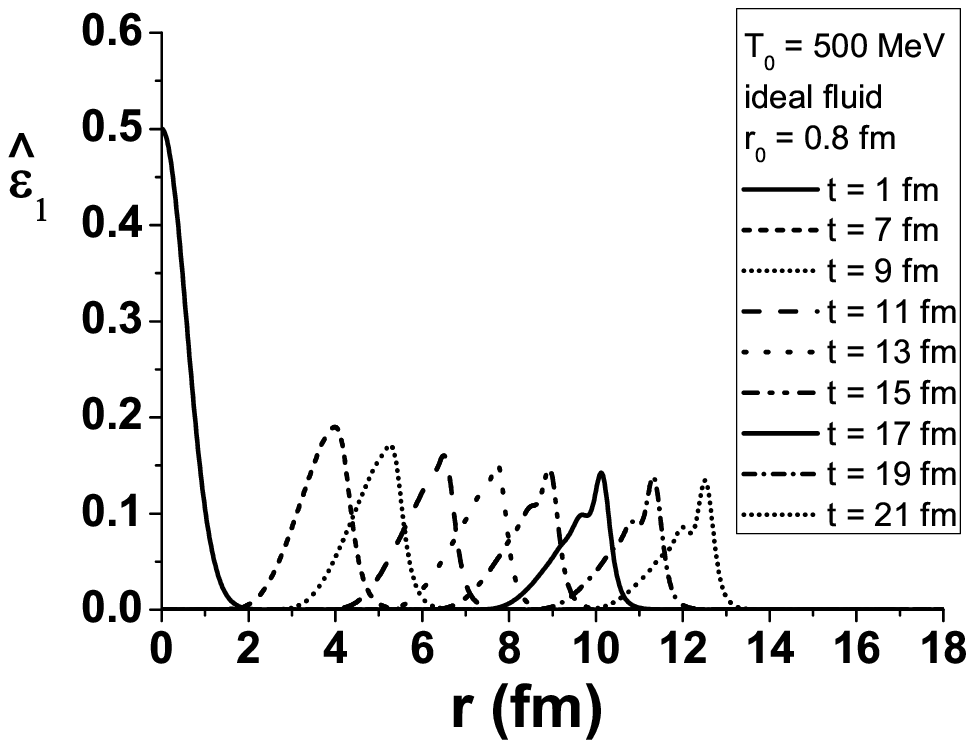}}
\subfigure[ ]{\label{fig:fourth}
\includegraphics[width=0.48\textwidth]{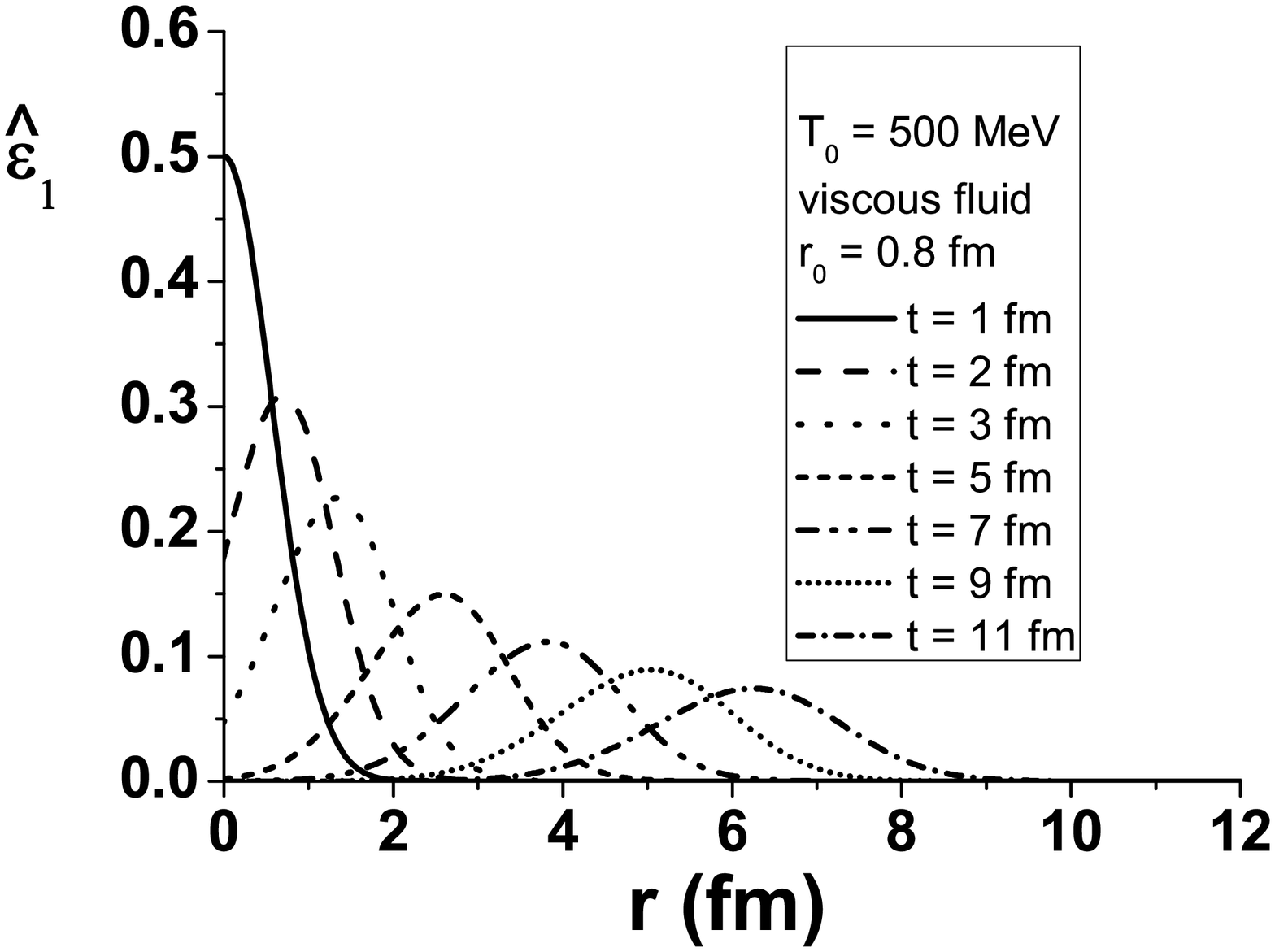}}
\end{center}
\caption{Numerical solutions of (\ref{weq}) for an ideal fluid (left) and of  (\ref{burgers_final}) for a viscous fluid
(right). The solutions are for several times and for $T_0=150$ MeV (upper panels) and  $T_0=500$ MeV (lower panels).
The initial amplitudes  $0.5$ and the width is $r_{0}=0.8 \,\, fm$.}
\label{fig3}
\end{figure}

\begin{figure}[ht!]
\begin{center}
\subfigure[ ]{\label{fig:first}
\includegraphics[width=0.48\textwidth]{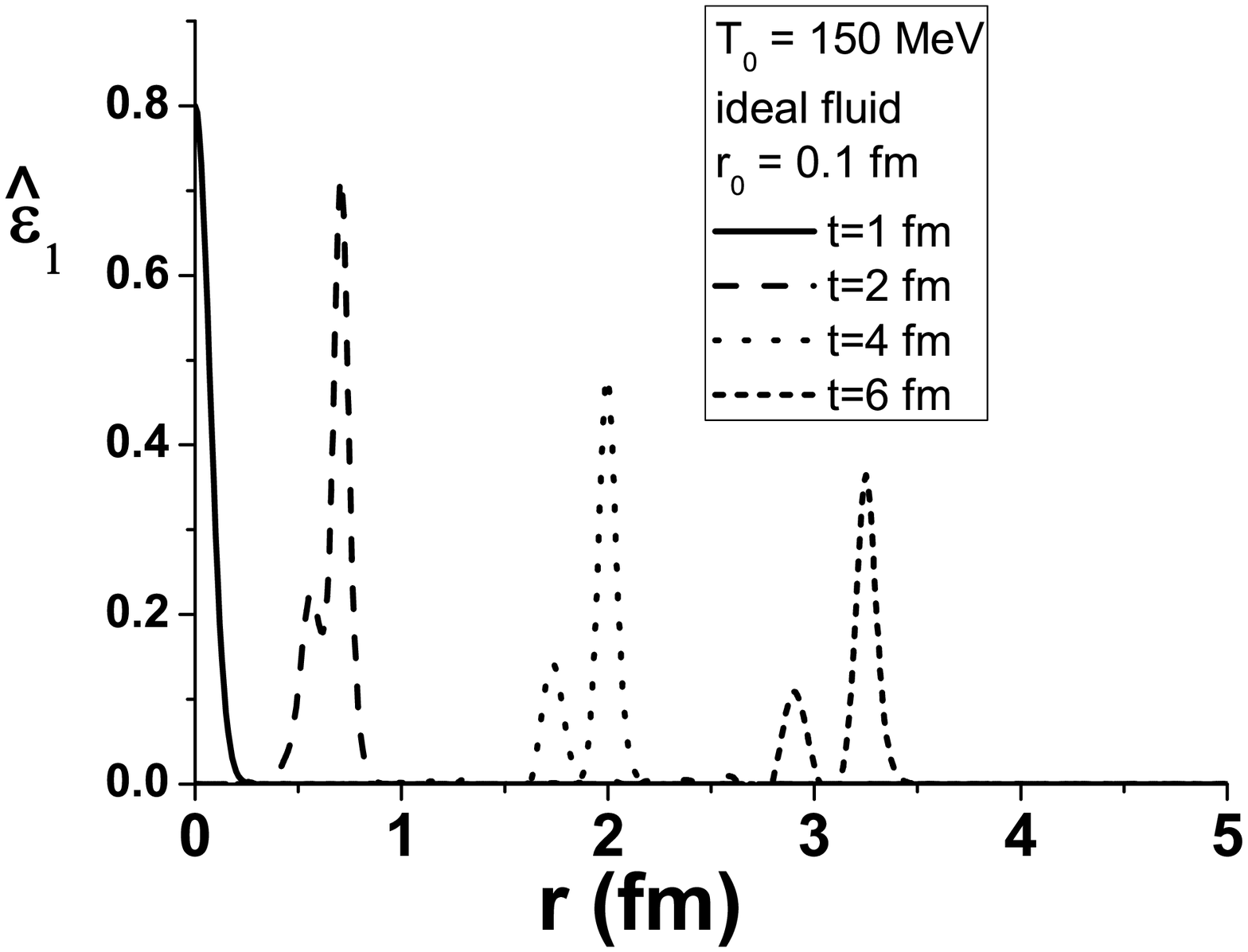}}
\subfigure[ ]{\label{fig:second}
\includegraphics[width=0.48\textwidth]{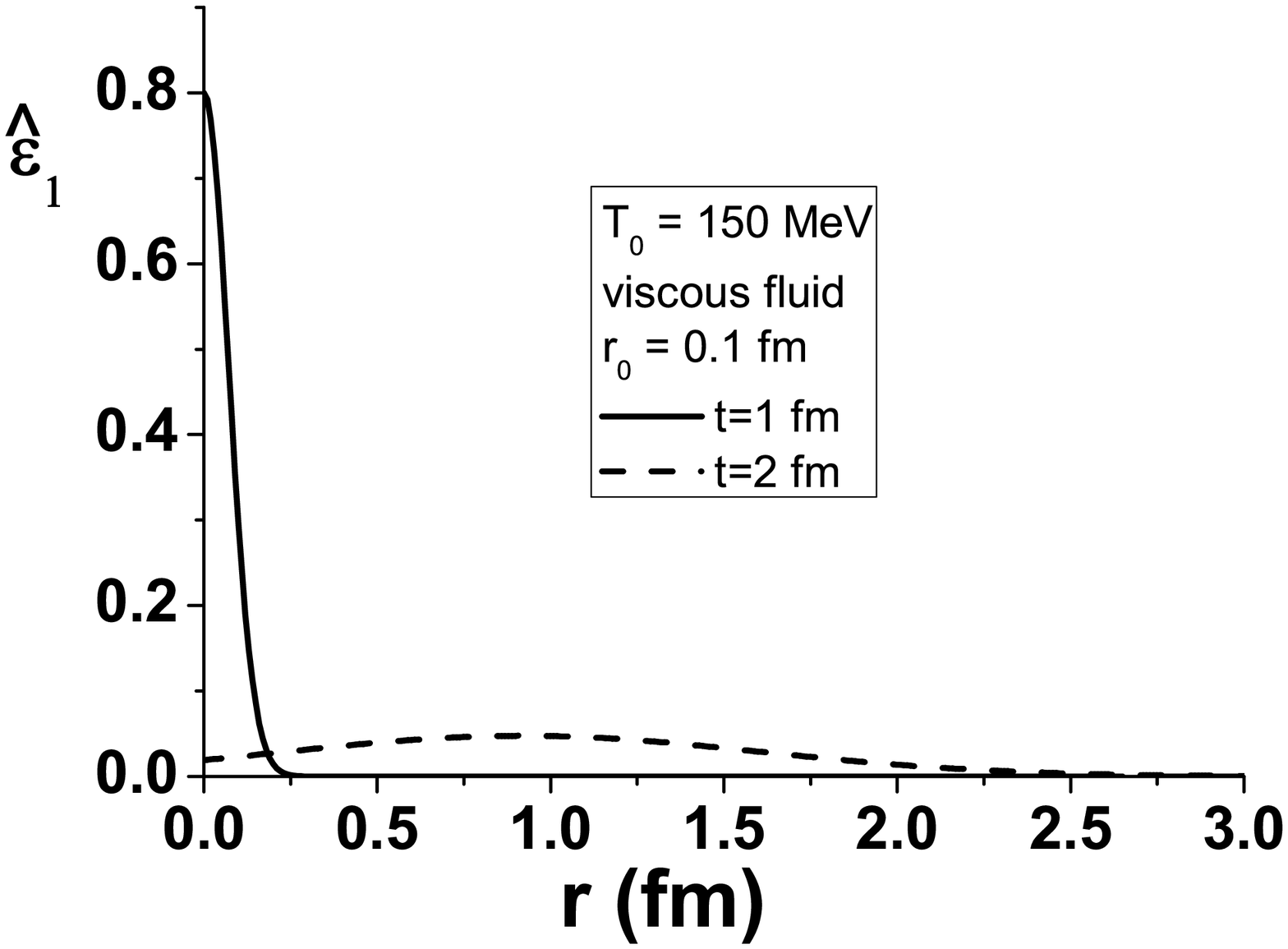}}\\
\subfigure[ ]{\label{fig:third}
\includegraphics[width=0.48\textwidth]{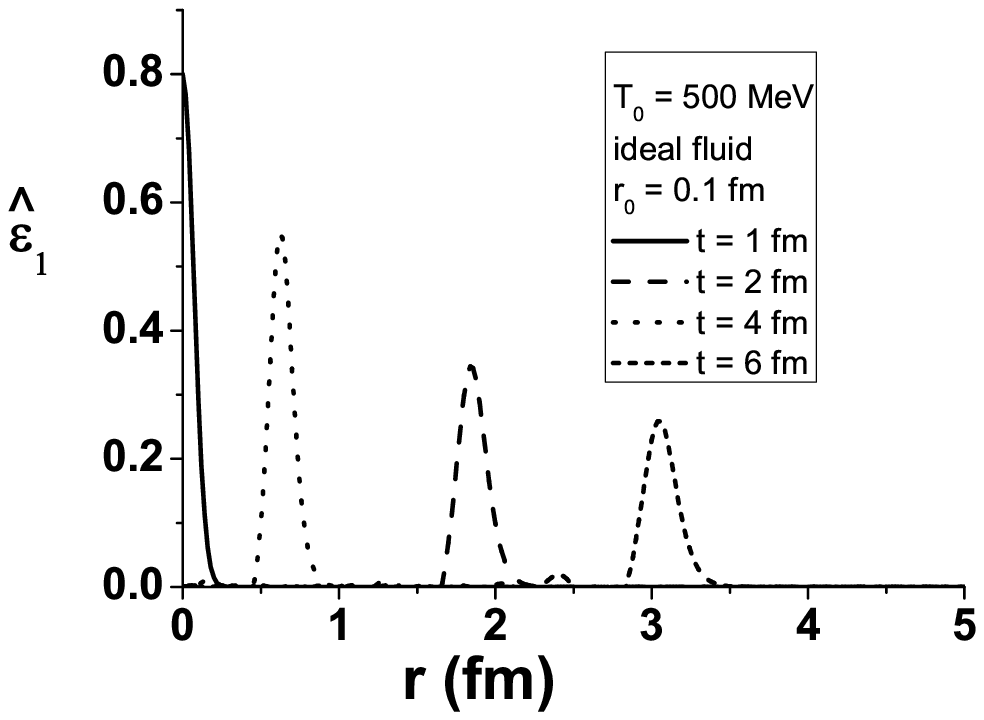}}
\subfigure[ ]{\label{fig:fourth}
\includegraphics[width=0.48\textwidth]{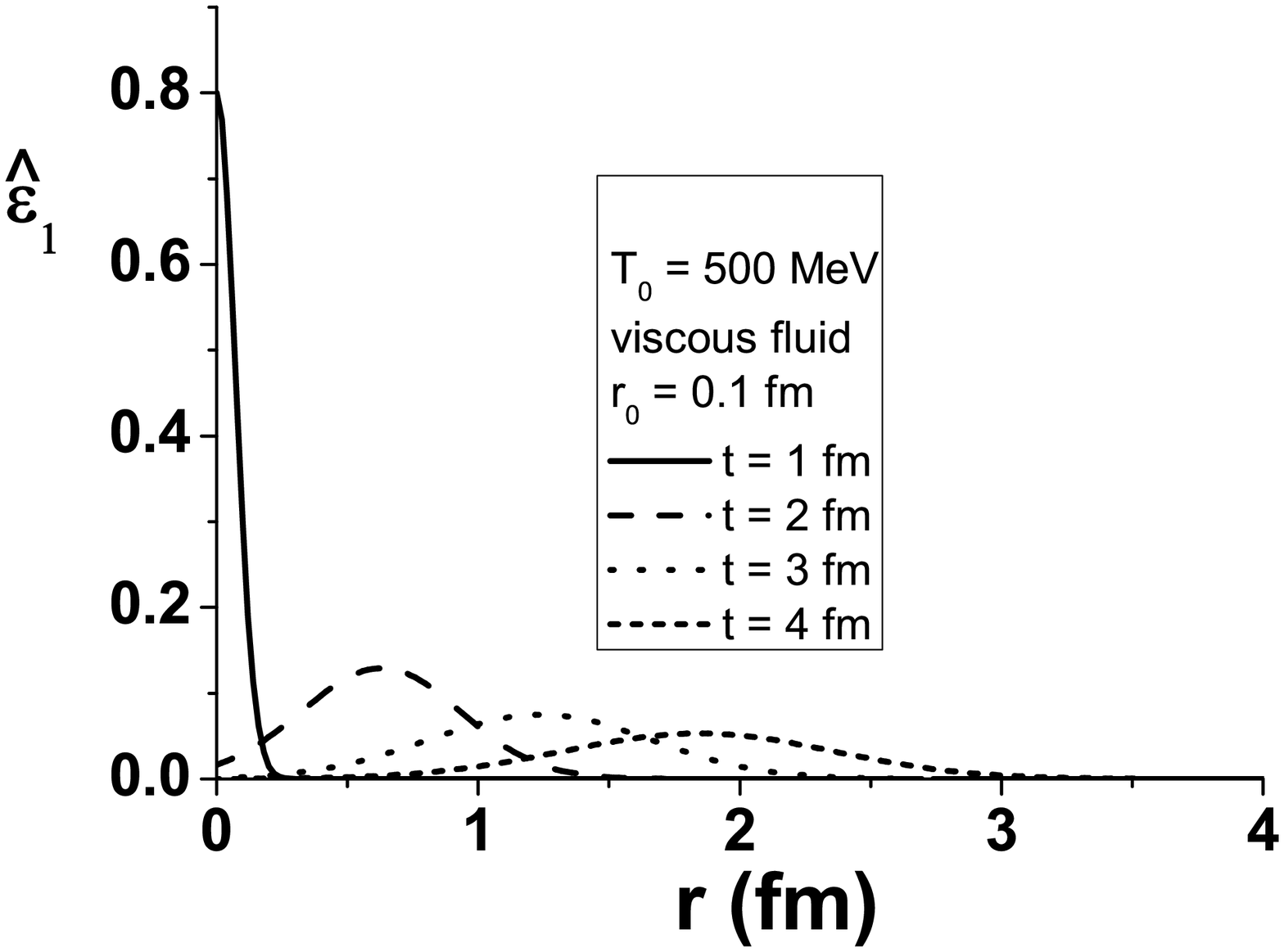}}
\end{center}
\caption{Numerical solutions of (\ref{weq}) for an ideal fluid (left) and of (\ref{burgers_final}) for a viscous fluid
(right). The solutions are for several times and for $T_0=150$ MeV (upper panels) and  $T_0=500$ MeV (lower panels).
The initial  amplitude is $0.8$ and the width is $r_{0}=0.1 \,\, fm$.}
\label{fig4}
\end{figure}

\begin{figure}[ht!]
\begin{center}
\subfigure[ ]{\label{fig:first}
\includegraphics[width=0.48\textwidth]{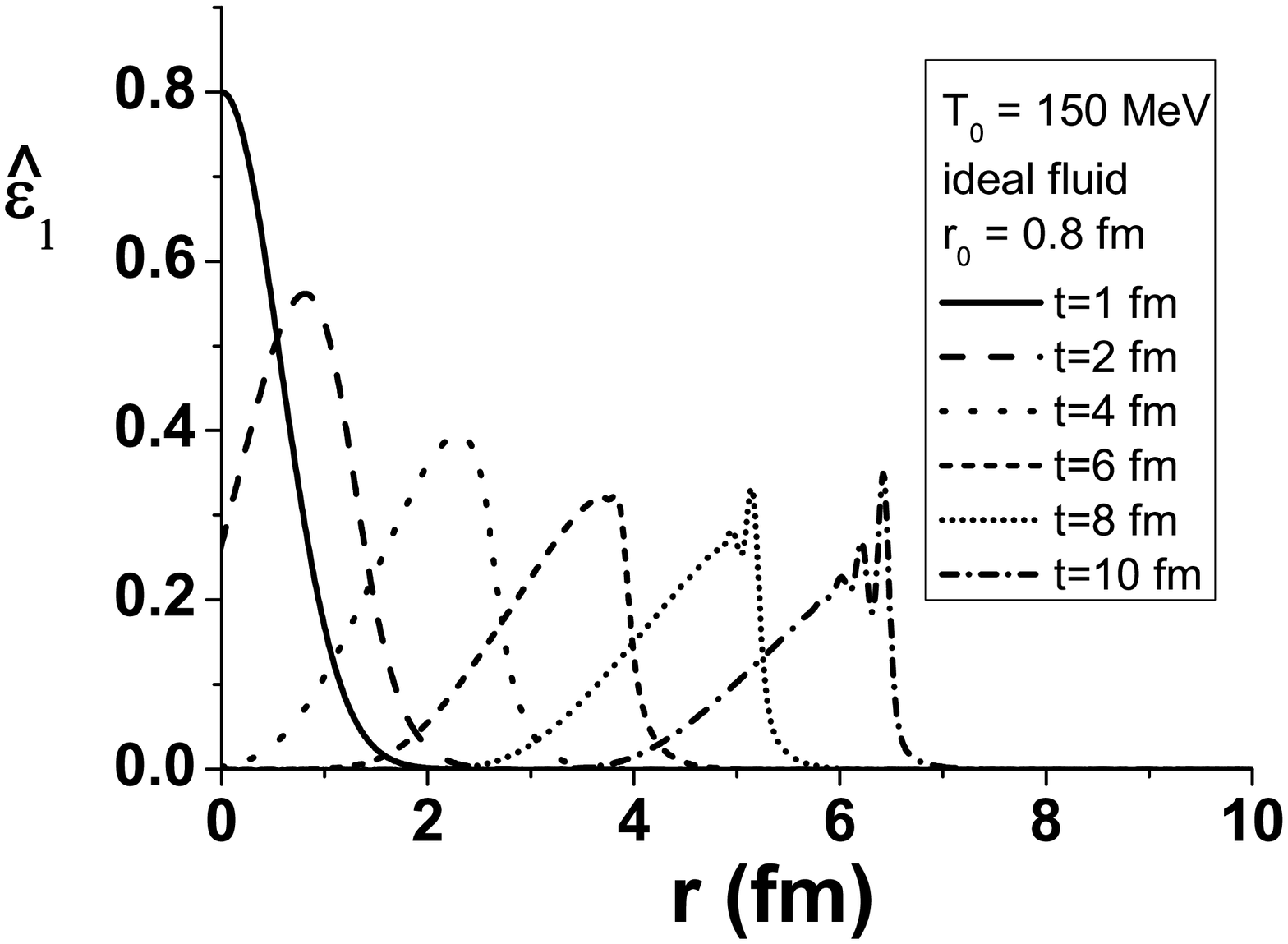}}
\subfigure[ ]{\label{fig:second}
\includegraphics[width=0.48\textwidth]{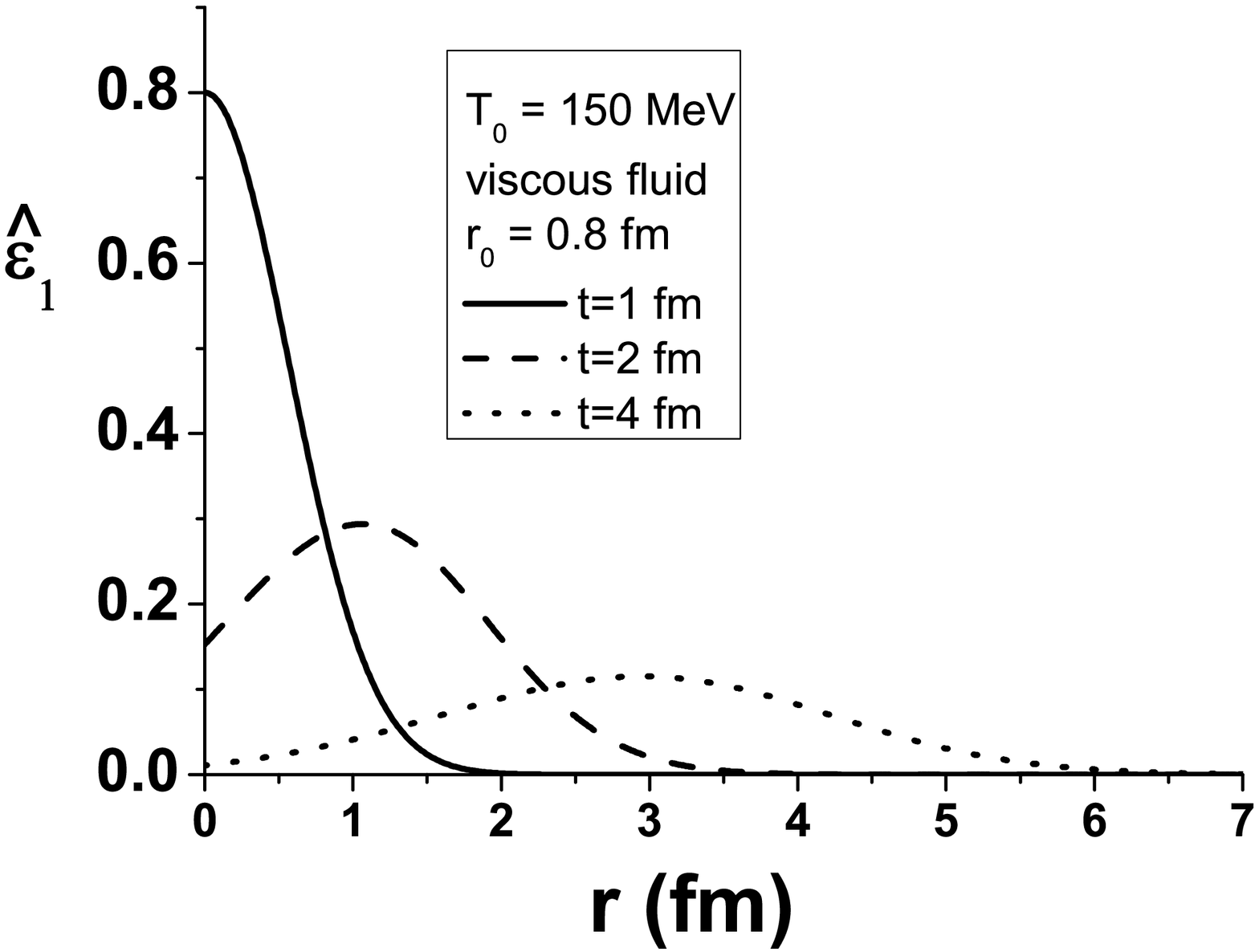}}\\
\subfigure[ ]{\label{fig:third}
\includegraphics[width=0.48\textwidth]{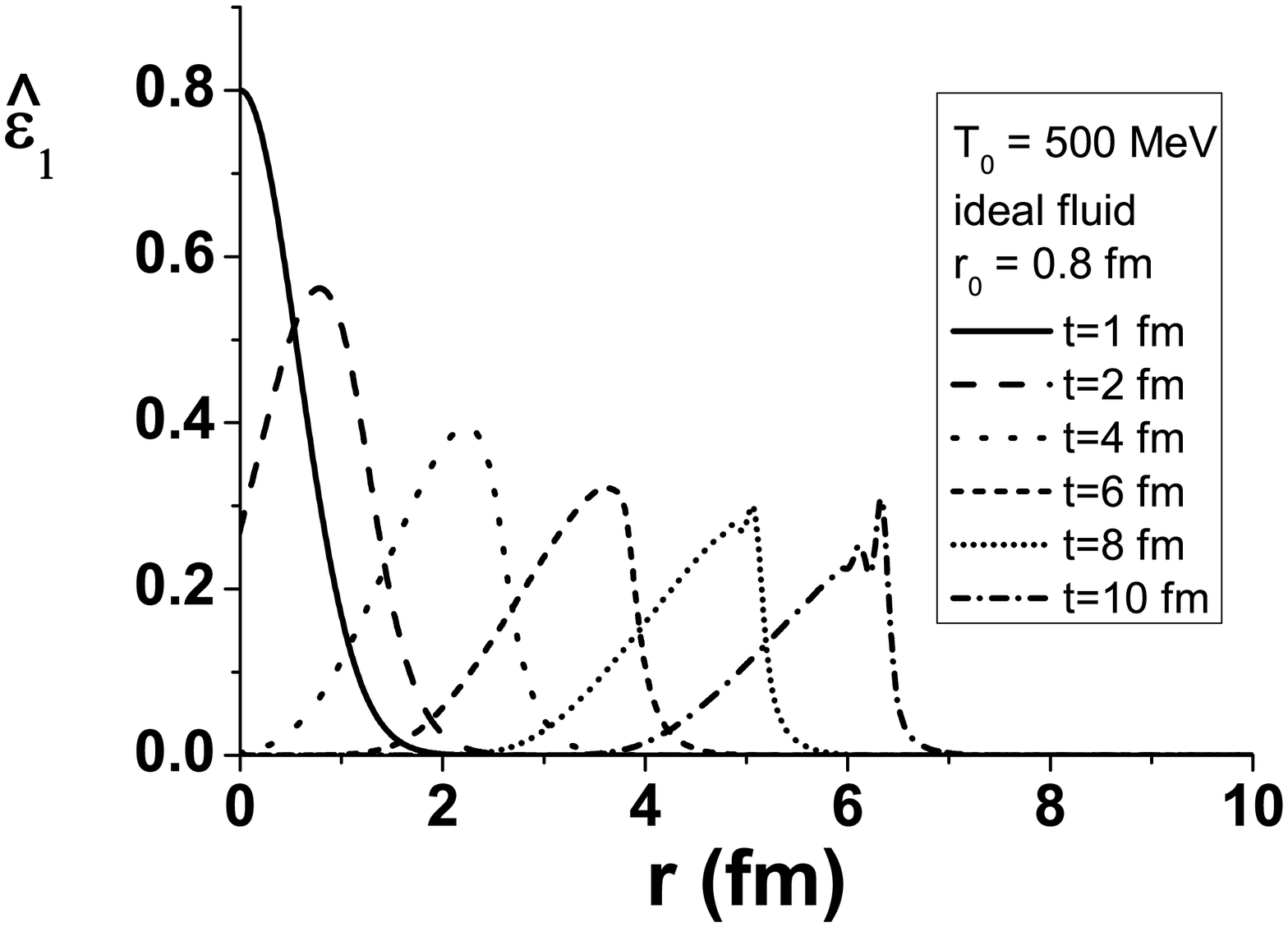}}
\subfigure[ ]{\label{fig:fourth}
\includegraphics[width=0.48\textwidth]{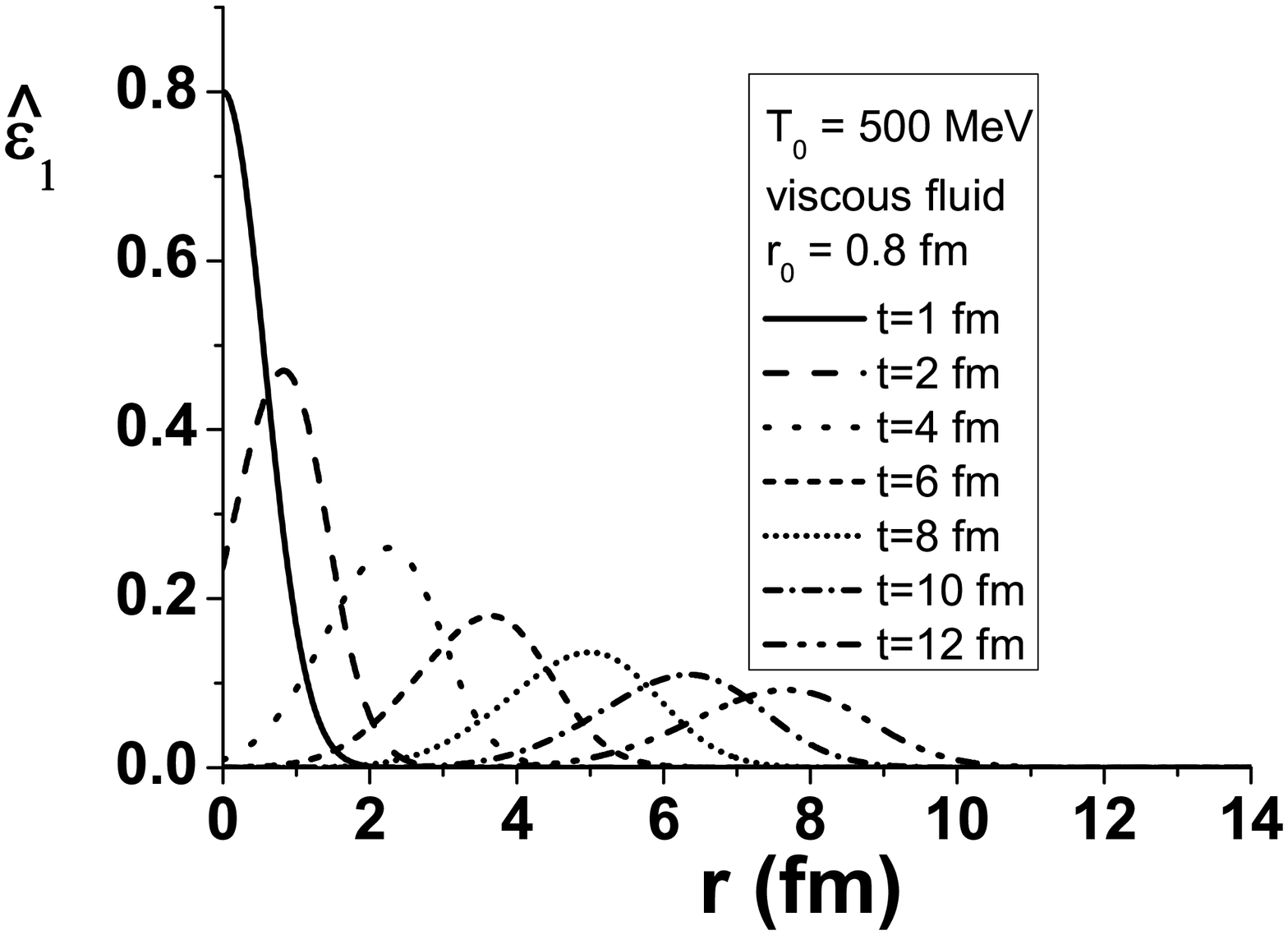}}
\end{center}
\caption{Numerical solutions of (\ref{weq}) for an ideal fluid (left) and of (\ref{burgers_final}) for a viscous fluid
(right). The solutions are for several times and for $T_0=150$ MeV (upper panels) and  $T_0=500$ MeV (lower panels).
The initial amplitude is $0.8$ and the width is $r_{0}=0.8 \,\, fm$.}
\label{fig5}
\end{figure}

\begin{figure}[ht!]
\begin{center}
\subfigure[ ]{\label{fig:first}
\includegraphics[width=0.48\textwidth]{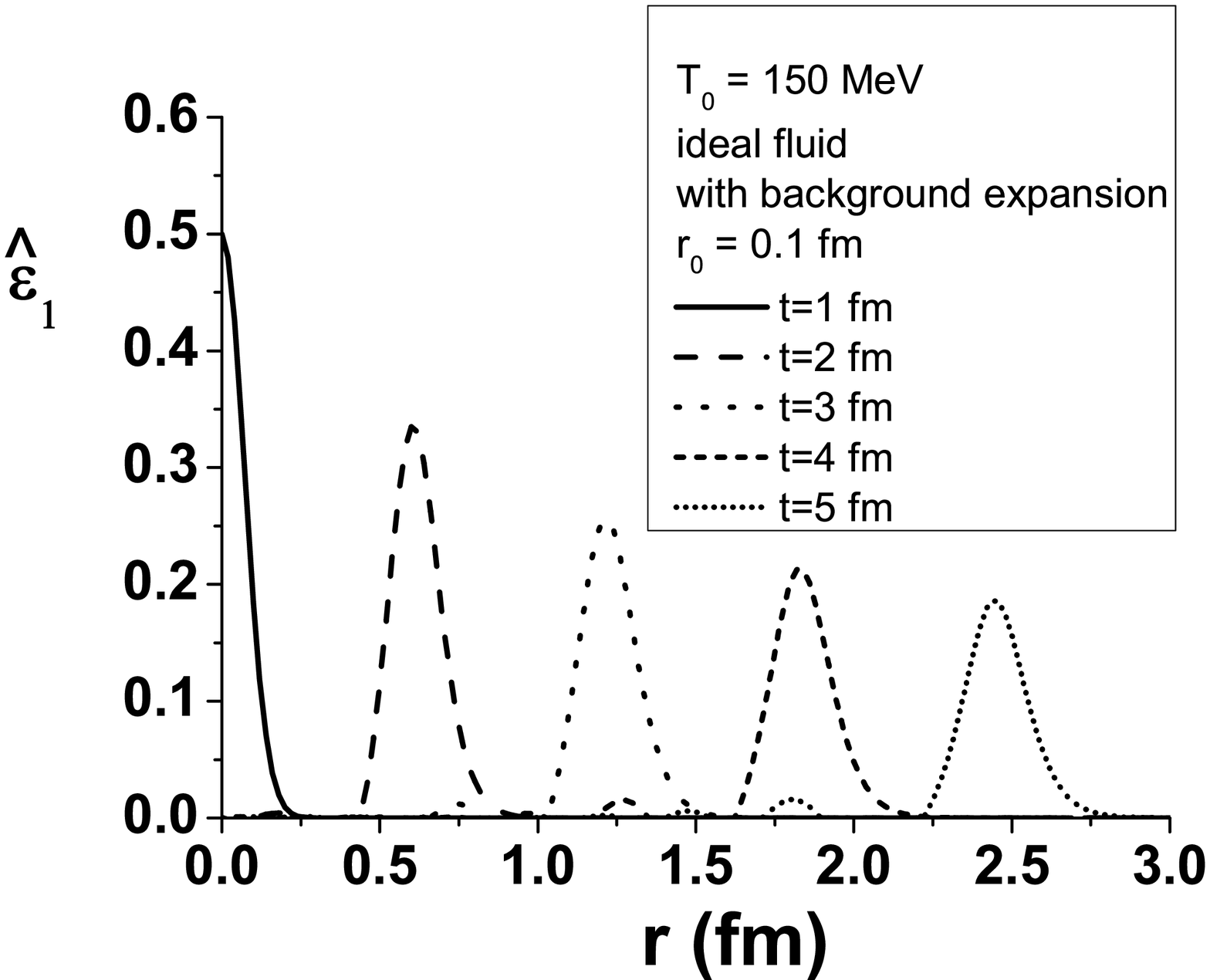}}
\subfigure[ ]{\label{fig:second}
\includegraphics[width=0.48\textwidth]{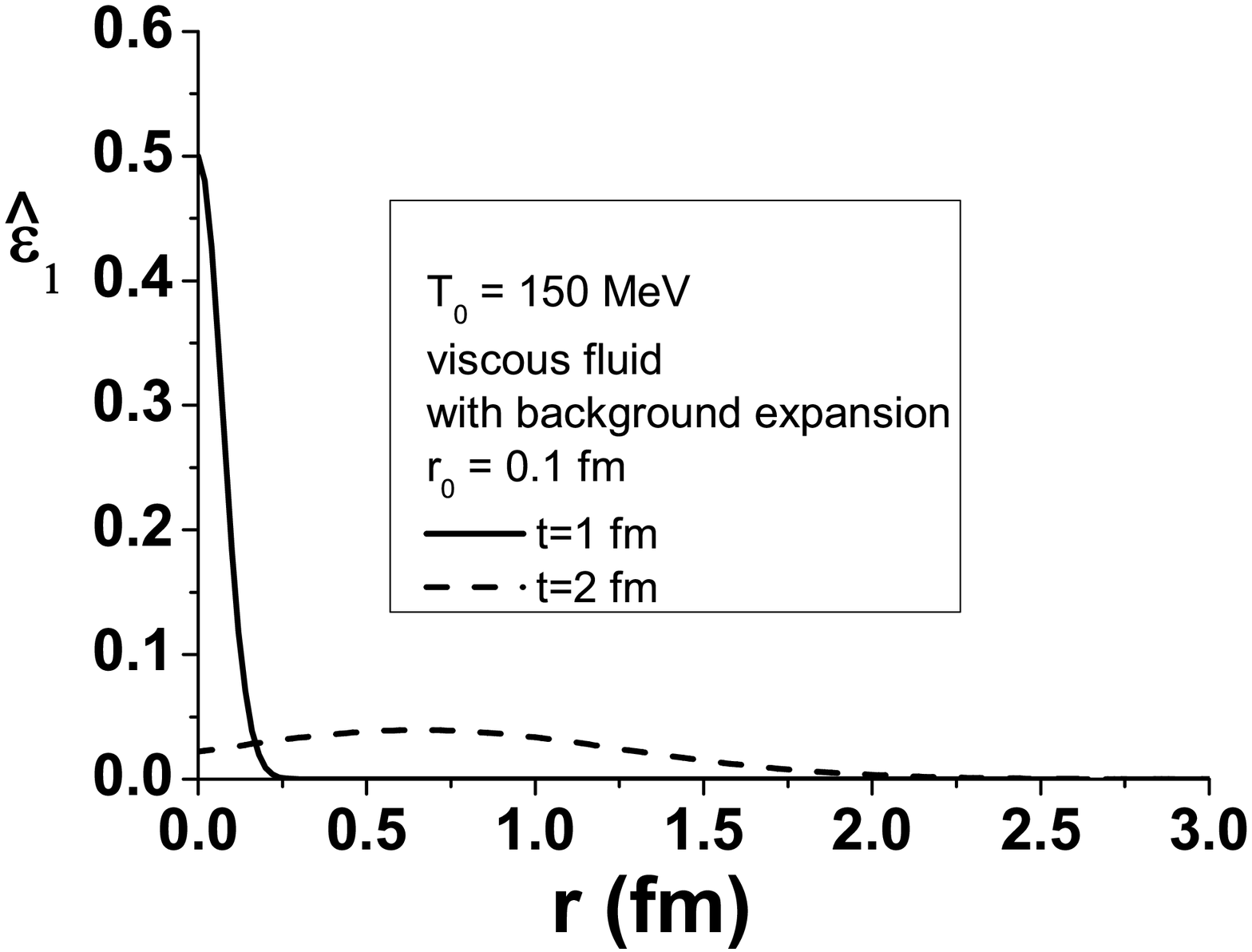}}\\
\subfigure[ ]{\label{fig:third}
\includegraphics[width=0.48\textwidth]{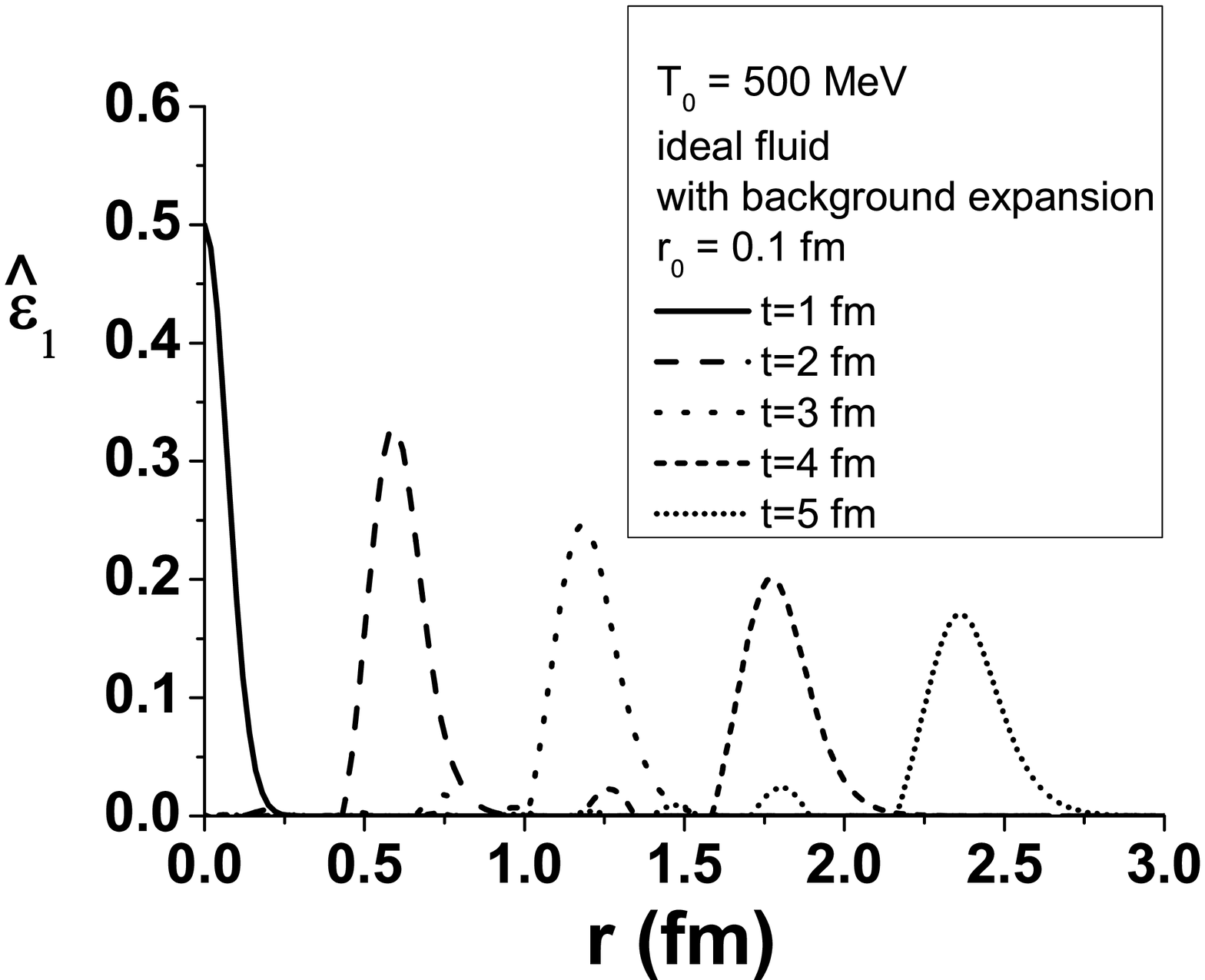}}
\subfigure[ ]{\label{fig:fourth}
\includegraphics[width=0.48\textwidth]{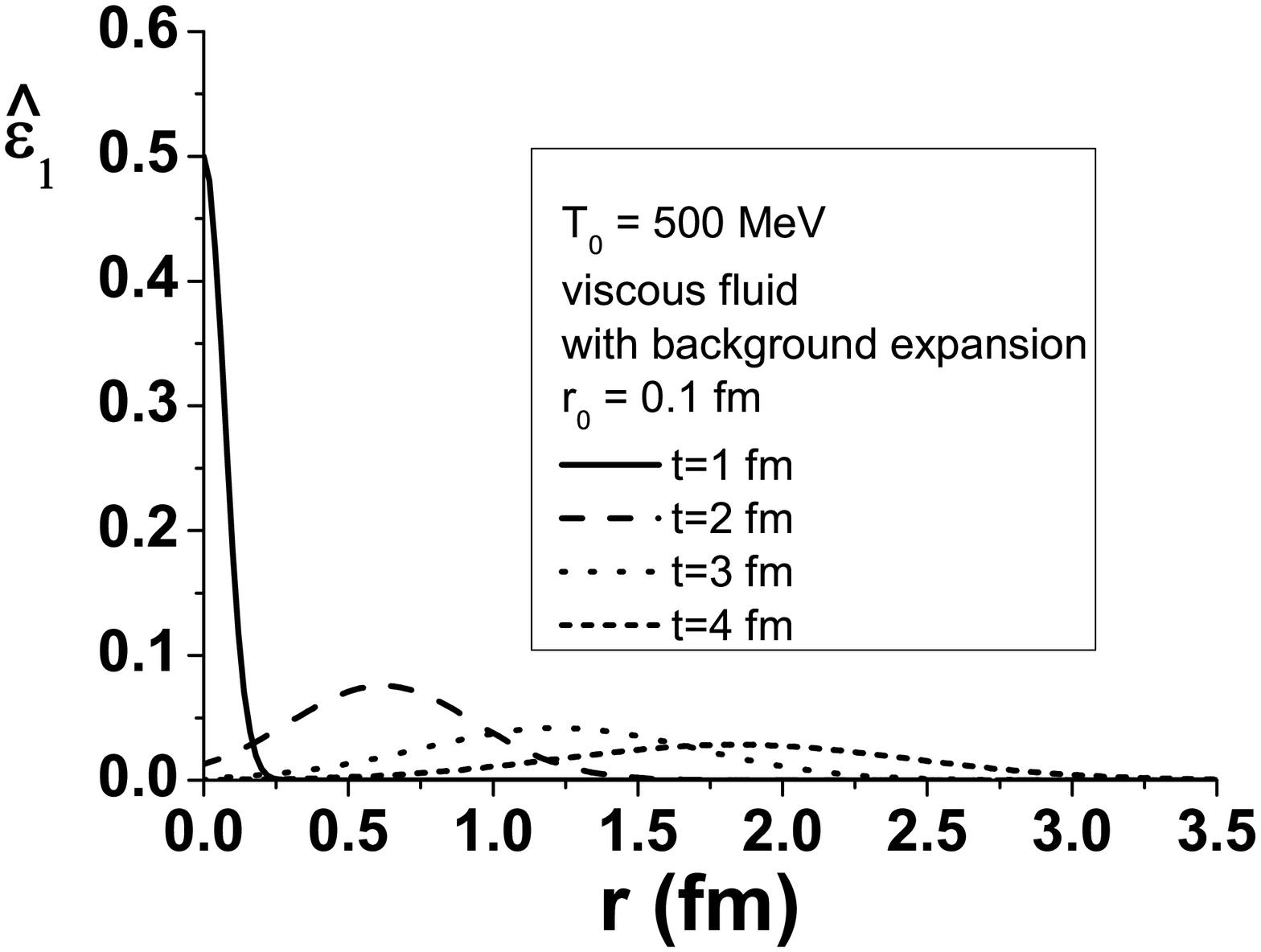}}
\end{center}
\caption{The same as  of Fig. \ref{fig2} with  background fluid expansion. The curves show numerical solutions of
(\ref{weq})  and  (\ref{burgers_final}) with a changing $T_0$, given by  Eq. (\ref{bjorkent}).}
\label{fig6}
\end{figure}

\begin{figure}[ht!]
\begin{center}
\subfigure[ ]{\label{fig:first}
\includegraphics[width=0.48\textwidth]{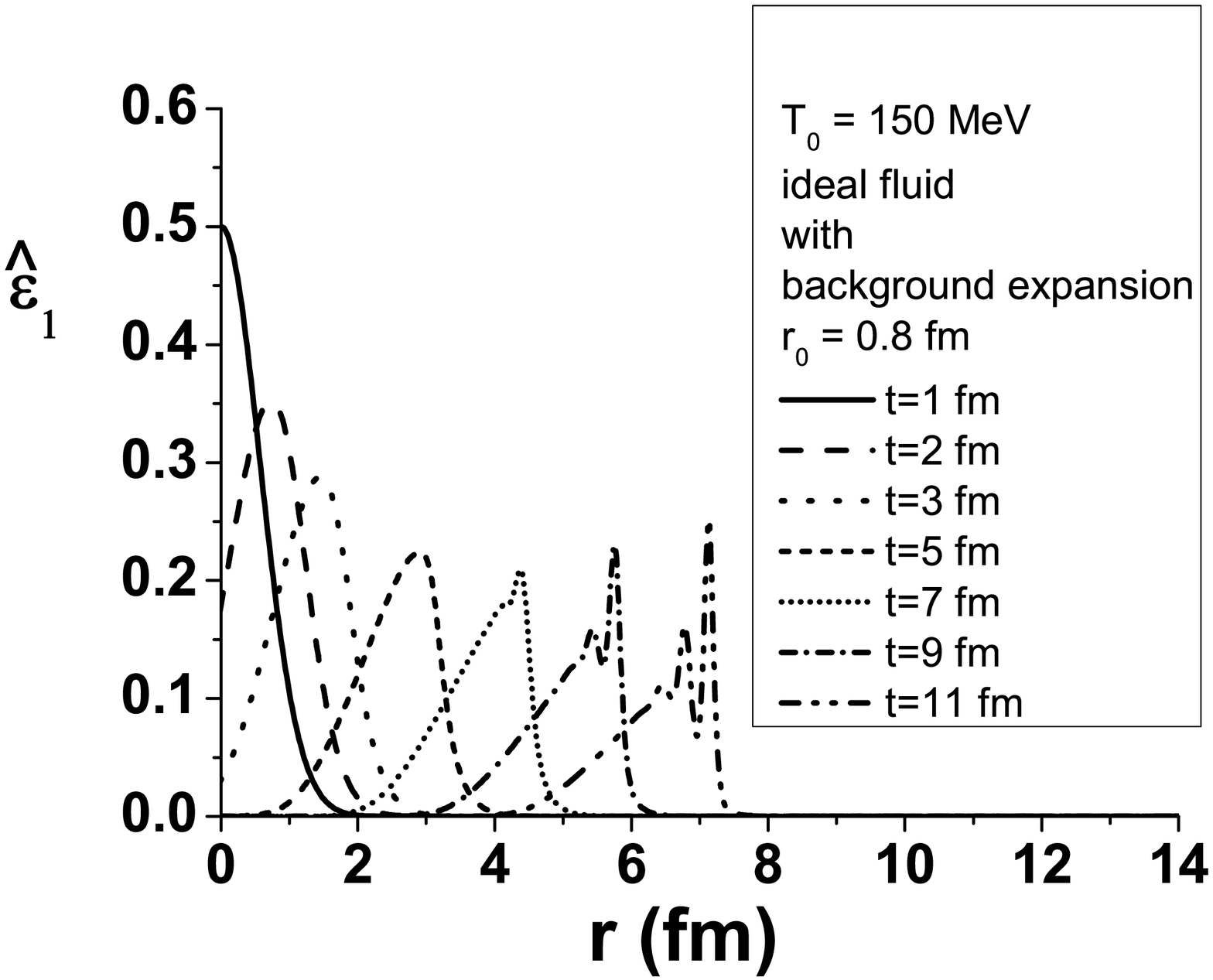}}
\subfigure[ ]{\label{fig:second}
\includegraphics[width=0.48\textwidth]{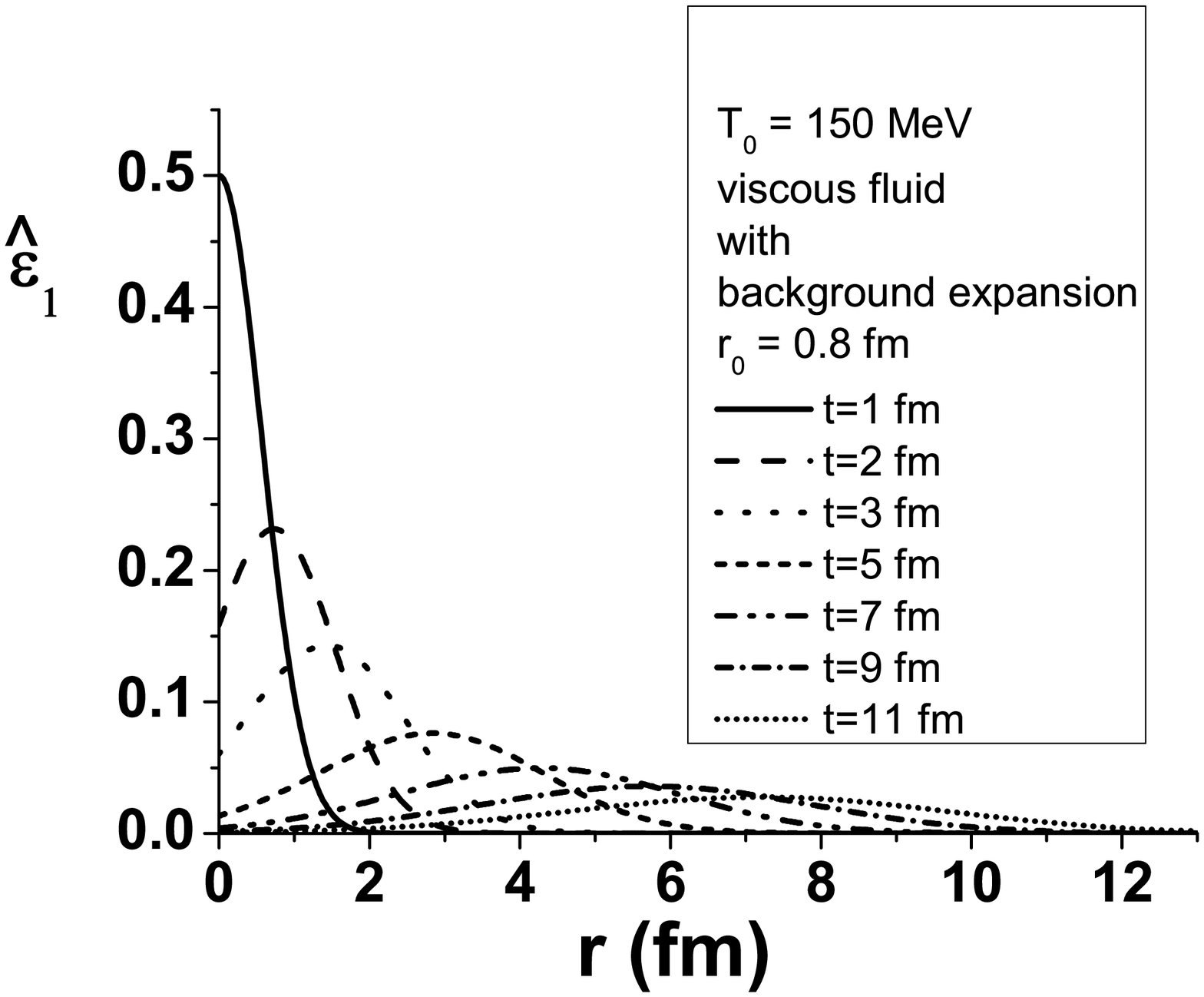}}\\
\subfigure[ ]{\label{fig:third}
\includegraphics[width=0.48\textwidth]{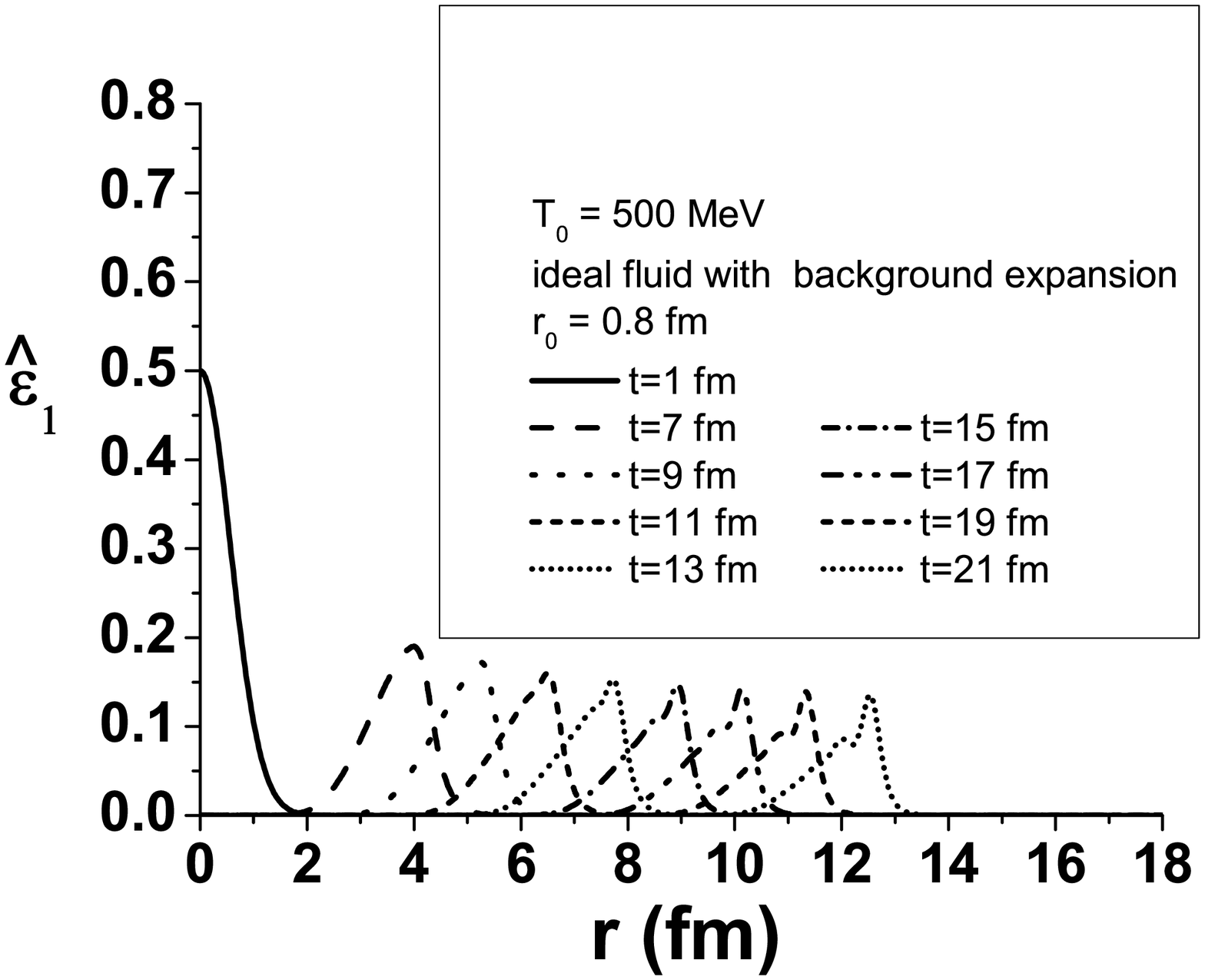}}
\subfigure[ ]{\label{fig:fourth}
\includegraphics[width=0.48\textwidth]{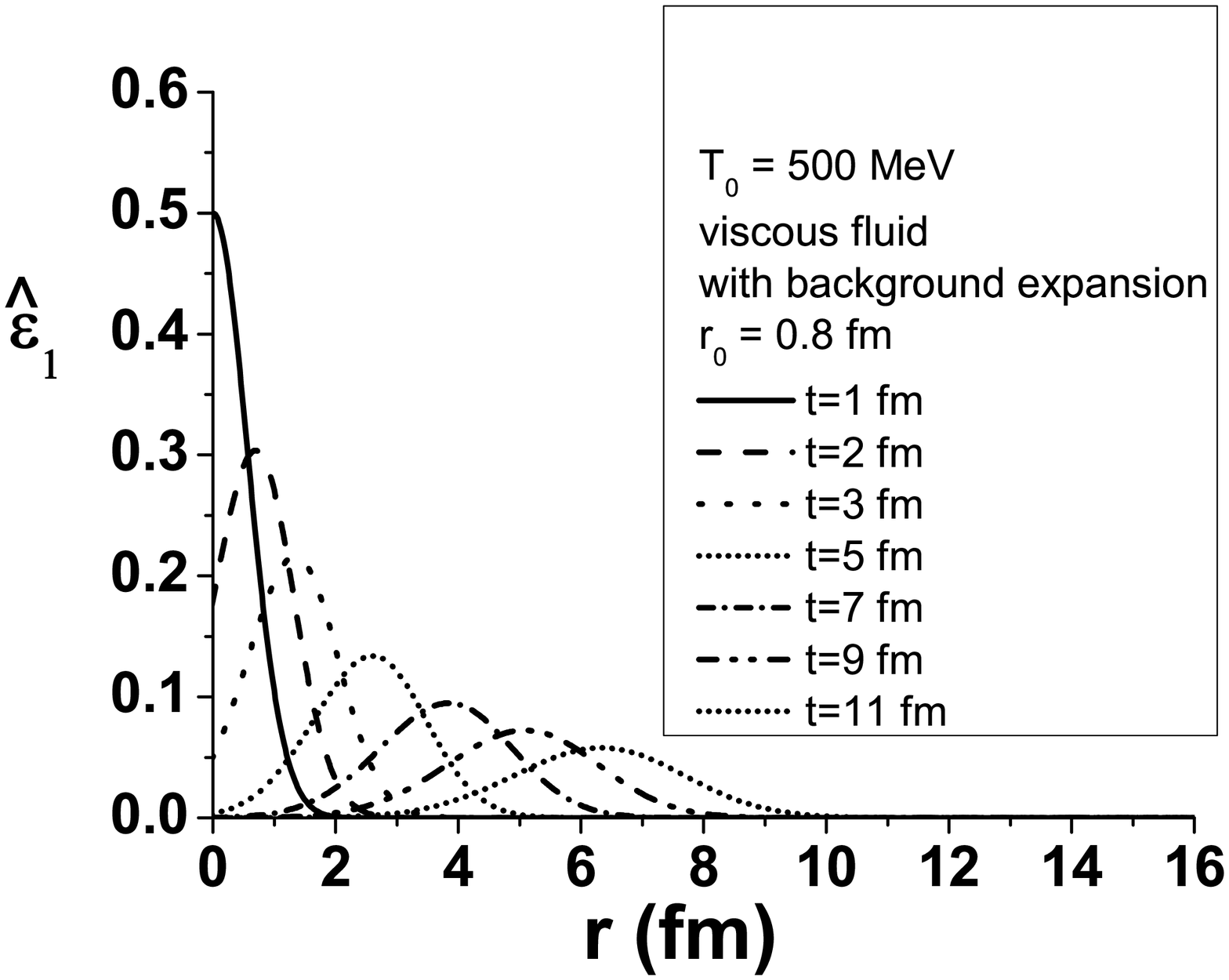}}
\end{center}
\caption{The same as  of Fig. \ref{fig3} with  background fluid expansion. The curves show numerical solutions of
(\ref{weq})  and  (\ref{burgers_final}) with a changing $T_0$, given by  Eq. (\ref{bjorkent}).}
\label{fig7}
\end{figure}

\begin{figure}[ht!]
\begin{center}
\subfigure[ ]{\label{fig:first}
\includegraphics[width=0.48\textwidth]{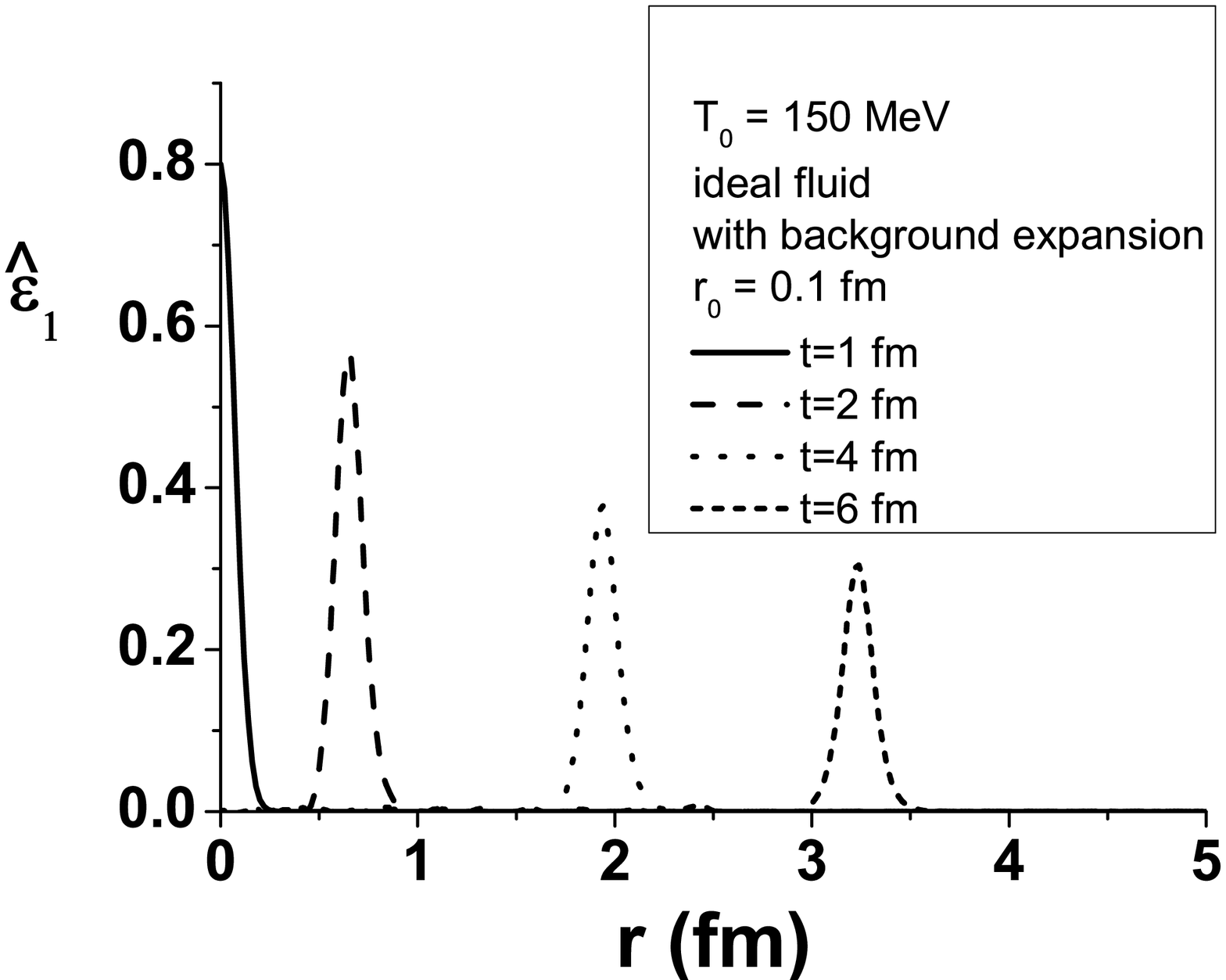}}
\subfigure[ ]{\label{fig:second}
\includegraphics[width=0.48\textwidth]{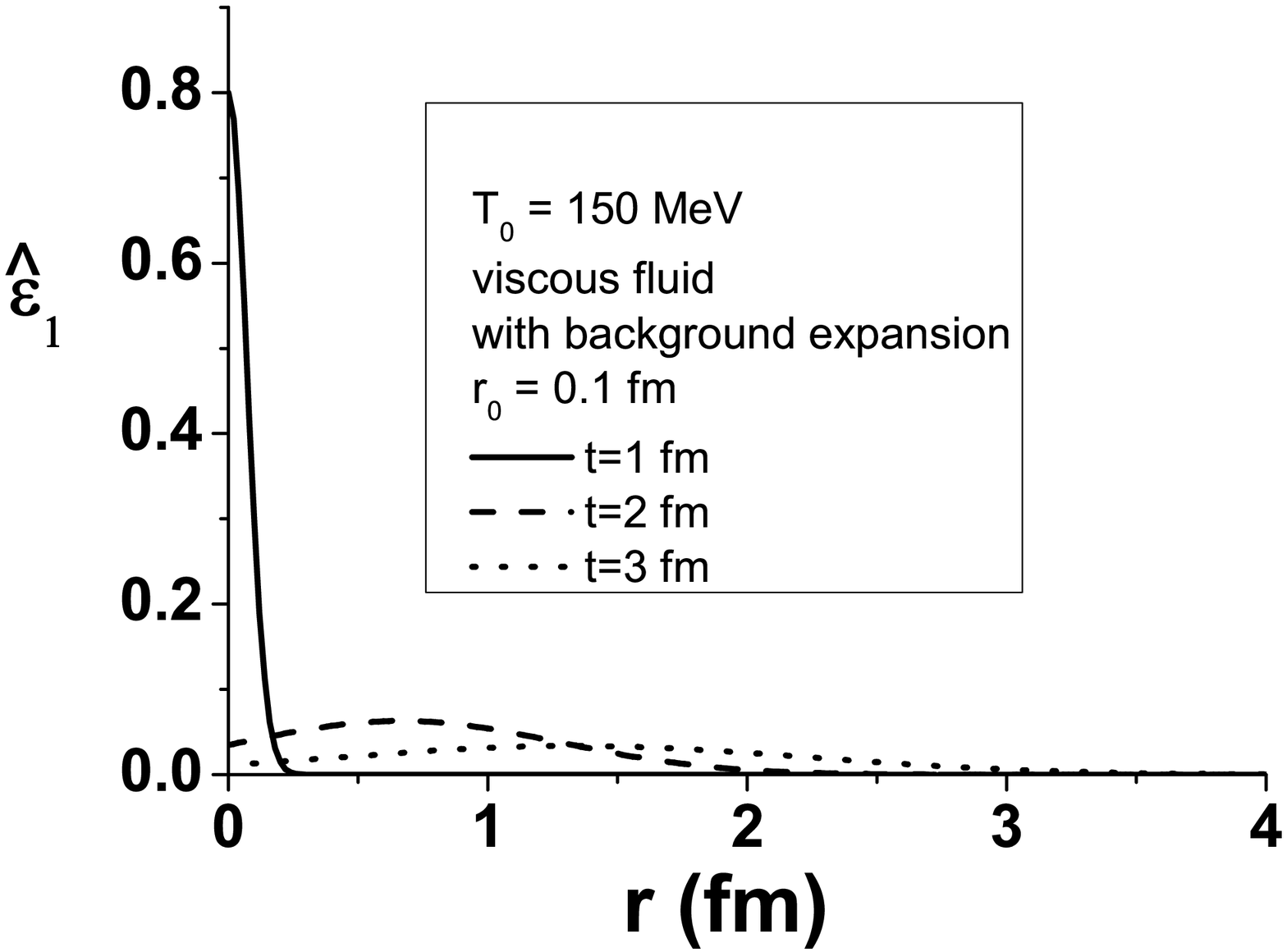}}\\
\subfigure[ ]{\label{fig:third}
\includegraphics[width=0.48\textwidth]{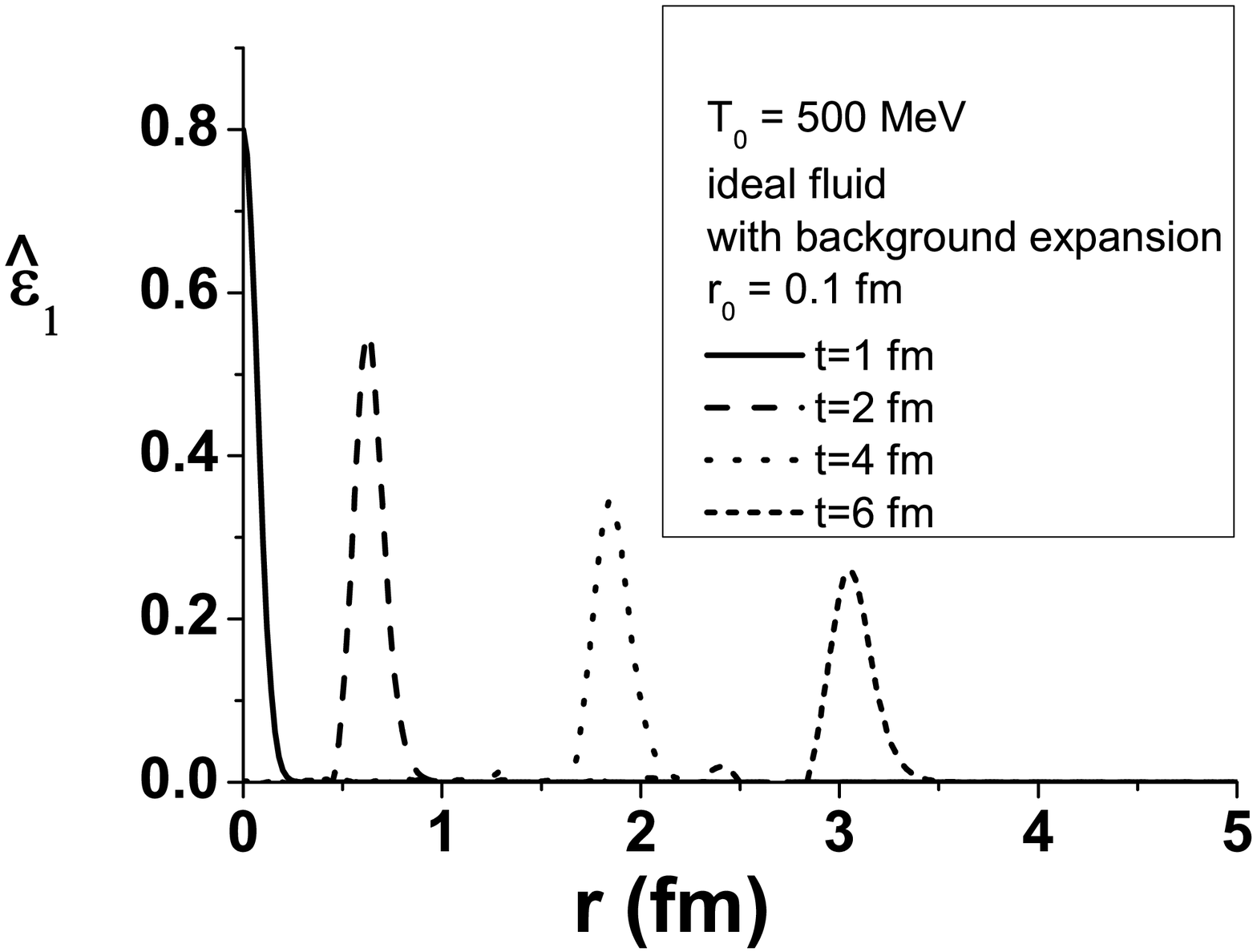}}
\subfigure[ ]{\label{fig:fourth}
\includegraphics[width=0.48\textwidth]{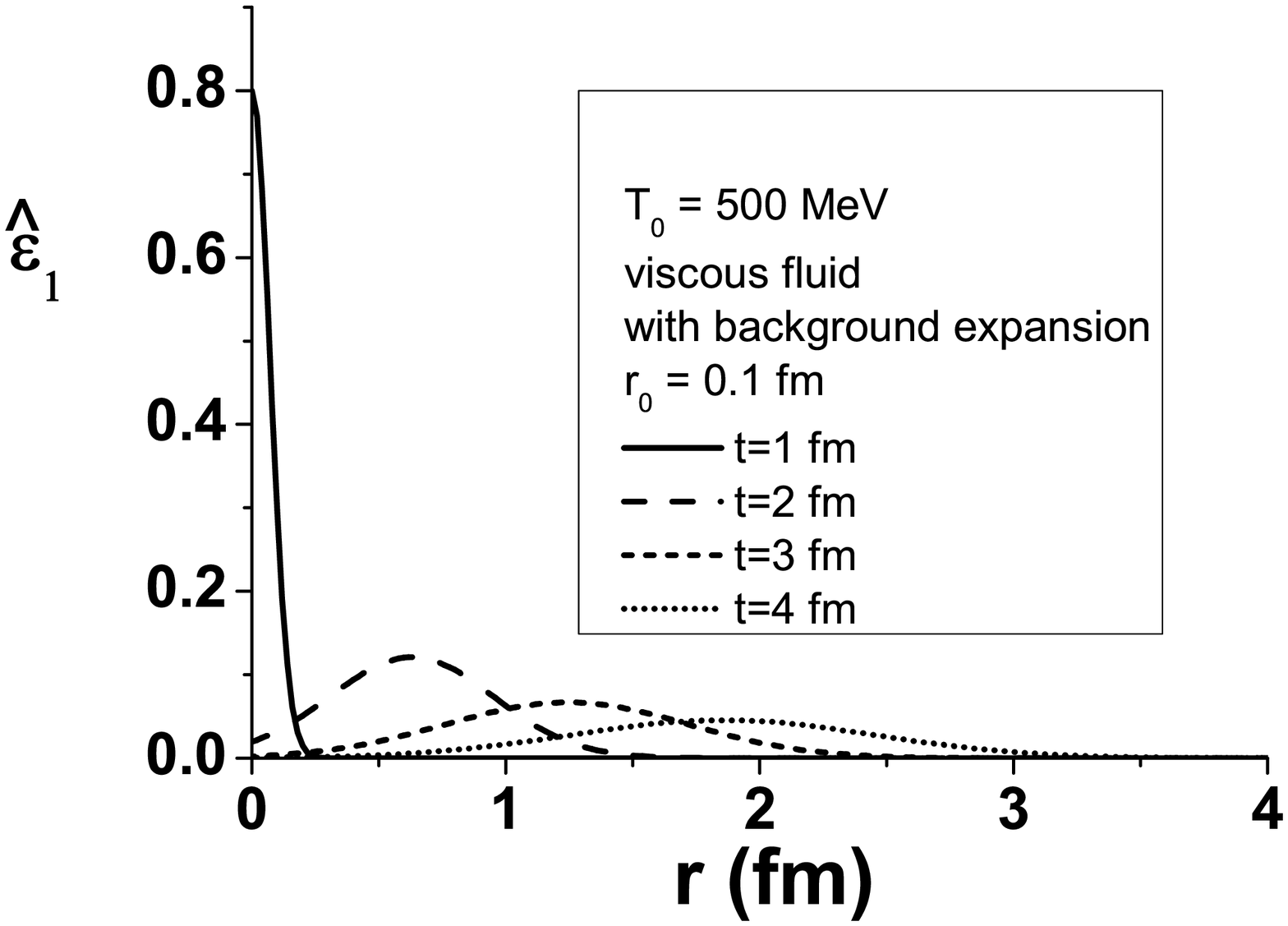}}
\end{center}
\caption{The same as  of Fig. \ref{fig4} with  background fluid expansion. The curves show numerical solutions of
(\ref{weq})  and  (\ref{burgers_final}) with a changing $T_0$, given by  Eq. (\ref{bjorkent}).}
\label{fig8}
\end{figure}

\begin{figure}[ht!]
\begin{center}
\subfigure[ ]{\label{fig:first}
\includegraphics[width=0.48\textwidth]{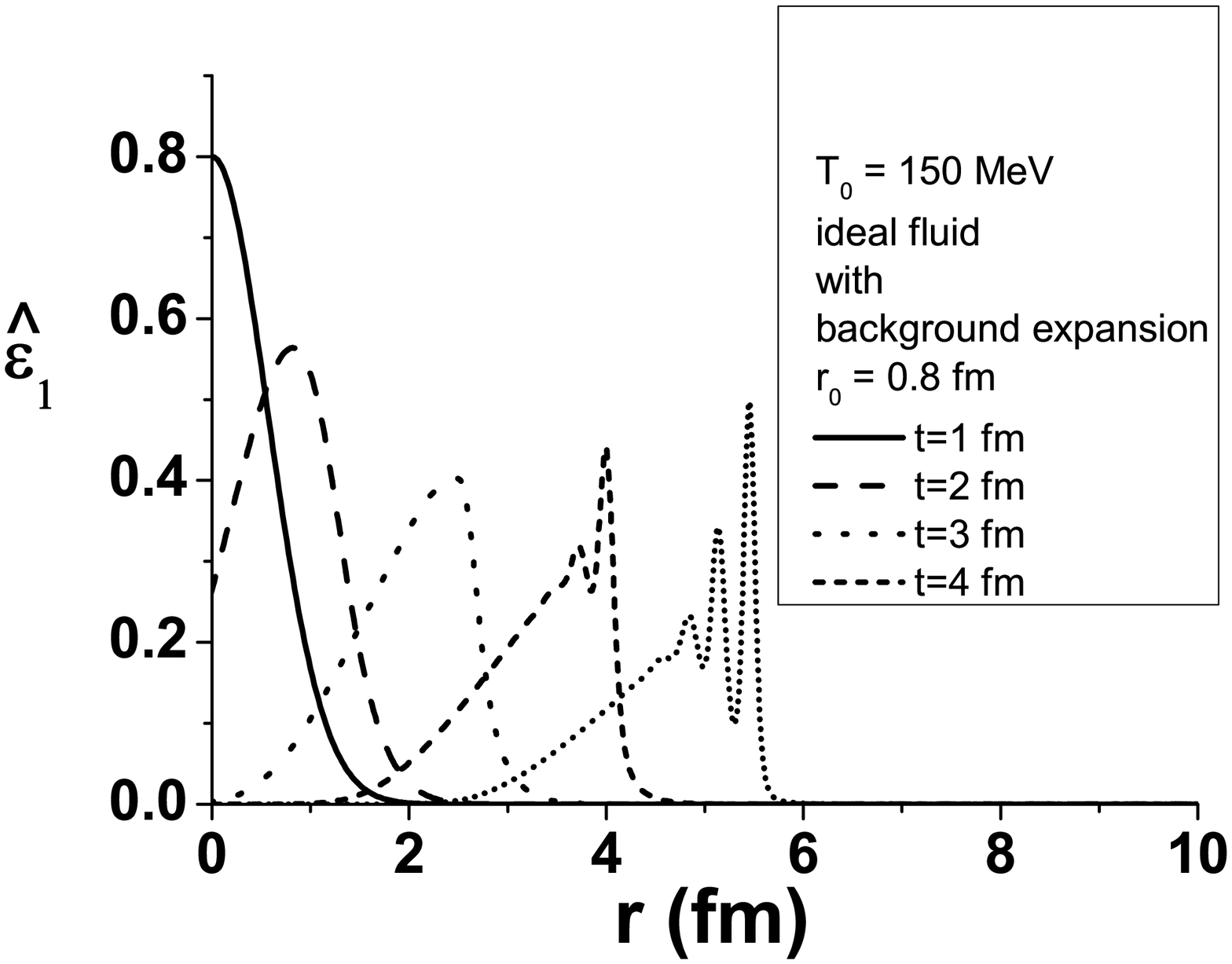}}
\subfigure[ ]{\label{fig:second}
\includegraphics[width=0.48\textwidth]{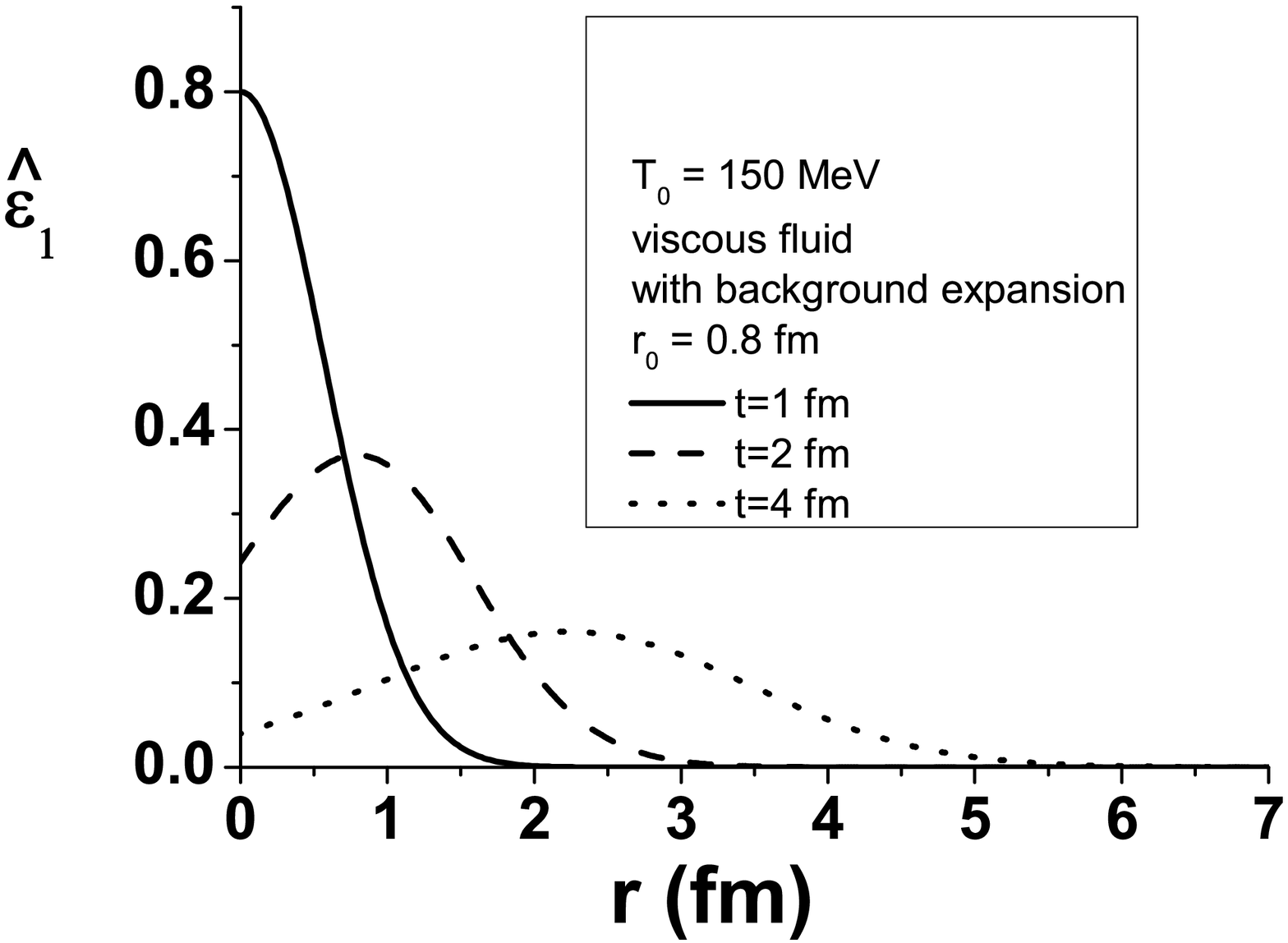}}\\
\subfigure[ ]{\label{fig:third}
\includegraphics[width=0.48\textwidth]{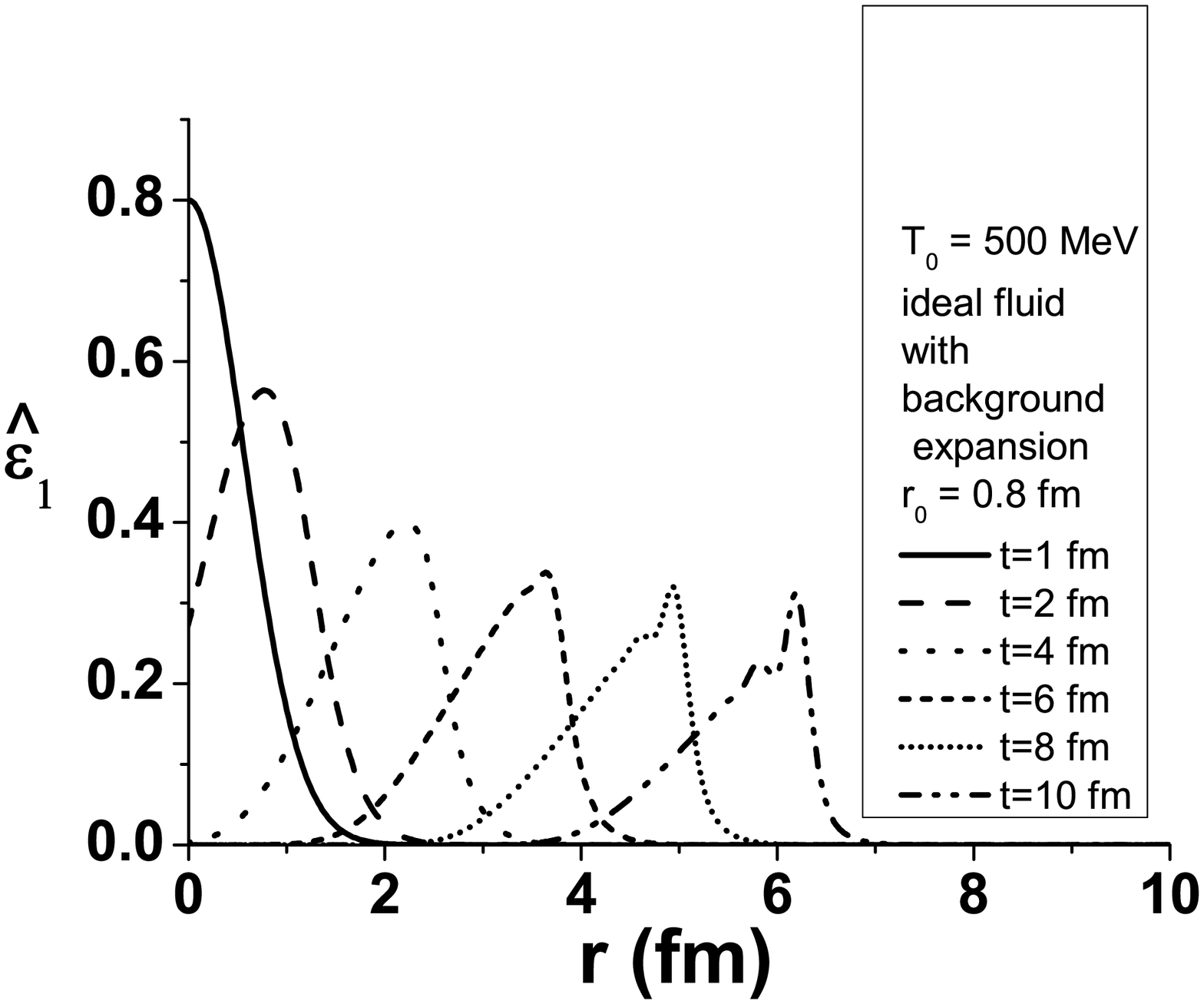}}
\subfigure[ ]{\label{fig:fourth}
\includegraphics[width=0.48\textwidth]{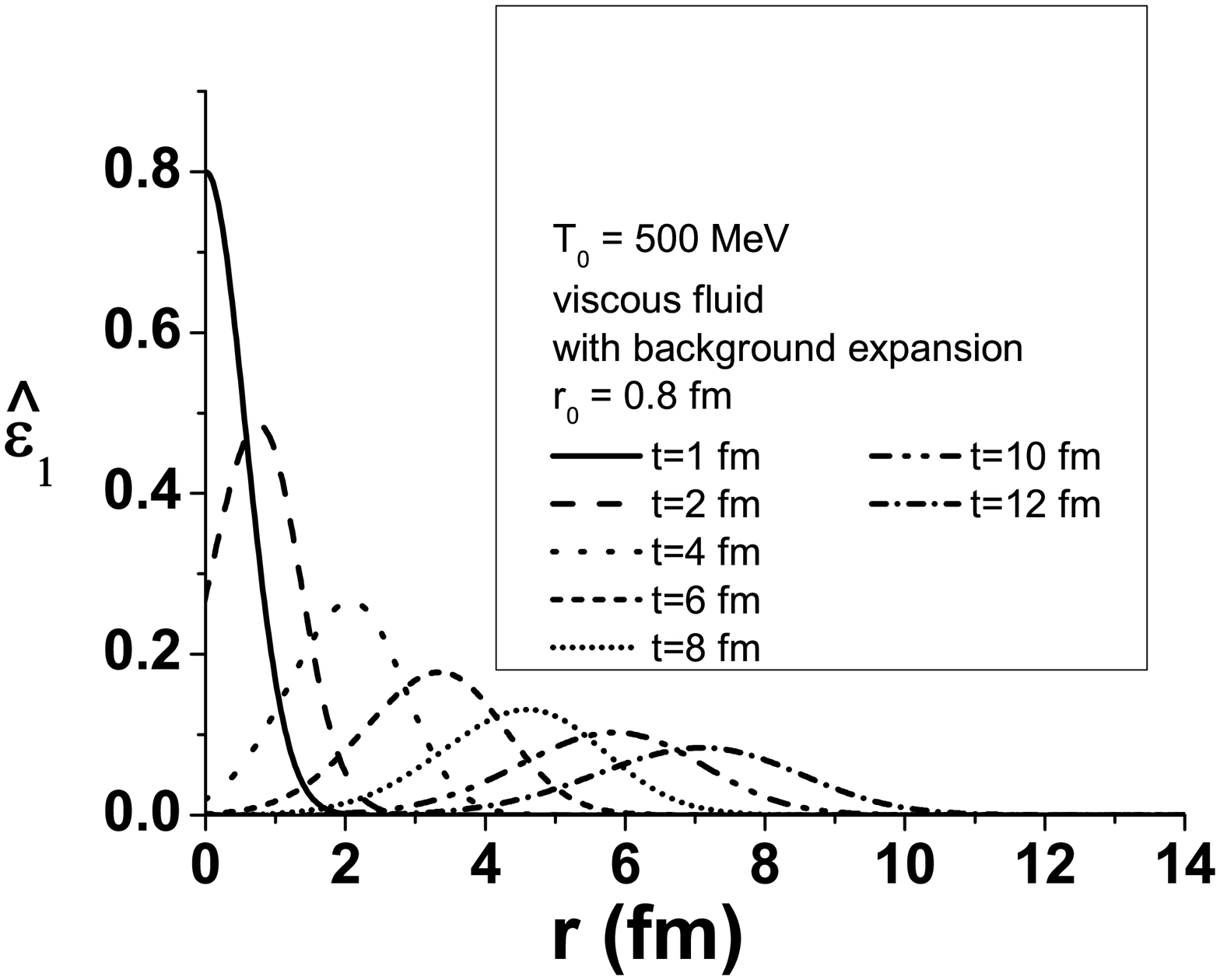}}
\end{center}
\caption{The same as  of Fig. \ref{fig5} with  background fluid expansion. The curves show numerical solutions of
(\ref{weq})  and  (\ref{burgers_final}) with a changing $T_0$, given by  Eq. (\ref{bjorkent}).}
\label{fig9}
\end{figure}

The most striking finding is the strong influence of viscosity. This can be seen in all figures and most clearly in the
comparison between   Fig. \ref{fig2}a)  and  \ref{fig2}b).  Viscosity  damps the amplitude of the pulse by a factor ten
in 1 fm! Increasing the temperature of the medium reduces the effect of viscosity as it can be seen from the factor $1/T_0$
in  (\ref{burgers_final}). However the comparison between  Fig. \ref{fig2}b) and \ref{fig2}d) shows that this reduction
is not very strong. The role played by viscosity is also reduced when the initial radius parameter of the tube goes from
$r_0=0.1$ to $0.8$ fm.  This is easy to understand looking at (\ref{condinit}) and then at the right hand side of
(\ref{burgers_final}). A broader initial distribution generates smaller spatial gradients appearing in the viscosity term
of  (\ref{burgers_final}), which becomes smaller. Nevertheless the attenuation of the initial pulse remains strong as compared
to the ideal fluid case. This situation is illustrated in Fig. \ref{fig3}, which is to be compared with Fig. \ref{fig2}. If we
increase the initial amplitude from $A=0.5$ to $A=0.8$, the relevance of viscosity remains the same. This can be checked by
comparing Fig. \ref{fig2} with Fig. \ref{fig4} and also comparing Fig. \ref{fig3} with Fig. \ref{fig5}. The introduction of the
background cooling preserves the difference between ideal and viscous fluids, as can be inferred from the comparison between
Figs. \ref{fig2}, \ref{fig3}, \ref{fig4}, \ref{fig5}  and their analogues with cooling Figs. \ref{fig6}, \ref{fig7}, \ref{fig8}
and  \ref{fig9}.

The introduction of viscosity in our calculations is what  more strongly changes them.
As anticipated  in the introduction, due to dissipation, viscosity strongly damps and broadens the tubes during their expansion and they
are  more easily mixed with the background fluid, loosing their influence on final state particle correlations.  This is a
robust conclusion of our numerical analysis since it remains valid in all situations considered.  Moreover viscosity  prevents
the perturbation wave from breaking, as can be seen comparing, for example, Fig. \ref{fig5}a) with \ref{fig5}b) or comparing
 Fig. \ref{fig7}a) with \ref{fig7}b).  Looking at the time evolution of the peaks of the pulses, we can have an idea of the
velocity with which they propagate. Comparing the left with right side of  all figures  we can see the velocity of the pulses is
only weakly changed by viscosity. This velocity is  defined by the sound velocity, which in our approach is the same both for
ideal and viscous fluids.

\subsubsection{Initial radius of the tube}

The solutions of nonlinear differential equations, such as (\ref{weq}) and (\ref{burgers_final}), are expected to depend on the
initial condition. We can check this dependence changing the parameters in (\ref{condinit}) and solving again both (\ref{weq}) and
(\ref{burgers_final}). The comparison between Fig. \ref{fig2} with Fig. \ref{fig3} and also the comparison
between Fig. \ref{fig4} with Fig. \ref{fig5} shows that, after viscosity, changes in the initial tube radius are those which
most substantially affect the tube evolution. Essentially thinner tube are more fragile. They break more easily, developing secondary
bumps (called ``radiation") and/or forming a wall with rapid oscillations at the edge. These instabilities may lead to loss of
localization and absorption of the tube by the medium.  Larger tubes, on the other hand, live longer in the plasma.
This conclusion may be relevant for the physics of particle correlation studied  now at RHIC and LHC. Moreover, the transverse size of
the tubes has physical origins. Glasma flux tubes are typically thinner \cite{raju12} than the tubes obtained in event generator based
on string models (see \cite{hamatube} and \cite{andrade} for details).

\subsubsection{Initial amplitude  of the tube}

Figs. \ref{fig4} and \ref{fig5} are repetitions of  \ref{fig2} and \ref{fig3} with a larger amplitude.   These cases are very
interesting for us because in  perturbations with larger amplitudes the  nonlinear effects become more important. The reductive
perturbation method (RPM) employed here is well suited to preserve the nonlinearities of the original equations and transfer them to
the equations which govern the evolution of perturbations.  From the figures we can conclude that pulses with larger amplitudes break
faster.  However the effect is not very pronounced because the range of variation considered here is relatively narrow:
$0.5 < A < 0.8$.

\subsubsection{Temperature}

Comparing in all figures the upper panels with the lower panels, we conclude that, in the case of ideal fluids,  there are only small
differences between them. This weak dependence on the temperature might have been anticipated from a closer
look at the coefficient of the nonlinear term in (\ref{weq}) and in (\ref{burgers_final}). The temperature dependent term in this
coefficient can vary only between zero and one, changing the overall coefficient at most by a factor two
(in realistic calculations the range of variation
is even narrower because  of the limits in the temperature: $150 < T_0 < 500$ MeV).  This weak sensitivity comes from all the
developments which led to (\ref{weq}) and is difficult to say what is more responsible for the final result, whether the equation of
state or the approximations adopted.  In contrast, the temperature dependence of the viscosity term in Eq.
(\ref{burgers_final}) is slightly stronger. Therefore the comparison of upper with lower panels on right side of all figures reveals
more pronounced  differences. In  viscous fluids the increase of  temperature decreases the amplitude of the pulse and makes it live
longer.  The second derivative term in the Burgers equations does not permit the breaking and dispersion of the pulse. However, together
with the geometrical term  $\hat\varepsilon_{1}/2t$, it causes the attenuation of the tube. In our calculations the viscosity
coefficients ($\eta$ and $\zeta$) were kept constant but they may be temperature dependent, enhancing the sensitivity of our results
to the temperature.

\subsubsection{Background expansion}

We have repeated all the calculations replacing (\ref{bwqcdTfinest}) and (\ref{bwqcdTfin_va}) by (\ref{weqbv}) and (\ref{weqbvi}).
The numerical solution of  the   latter equations  (neglecting the derivatives with respect to $z$)
is presented in Figs.  6 to  9. The effect of the Bjorken  cooling  is to slightly reduce the amplitude of the pulse, which can be
best seen comparing Fig. 2 with Fig. 6 and Fig. 3 with Fig. 7.  It is a small effect and this is very interesting. The cooling studied
here is a crude representation of the real three dimensional background fluid expansion. If we had found that cooling is important,
this would suggest that a realistic treatment of the background expansion would change completely the conclusions obtained so far.
This seems not to be the case.

\subsection{Final remarks}

If, on one hand, the similarity between the figures is somewhat deceptive (because of the weak dependence on the dynamical
ingredients), on the other hand they deliver a strong message: the tube
expands radially with a supersonic velocity and  in less than 4 fm/c  it becomes a ``ring'', with a hole  in the middle.
Moreover, by this time the amplitude is already reduced by a factor two and the tube (or ring) looses the strength to ``push away''
the surrounding matter. This agrees with the results found in \cite{andrade}, where the evolution of a tube was studied in a
different way. In that work the numerical solution of the hydrodynamical equations of the total system (tubular perturbation + background)
was obtained, whereas  here we have  isolated the perturbation from the background and written a differential equation for it.
We have provided an independent check of the results found in \cite{andrade} with the use of a different equation of state.

An important conclusion of our work is that viscosity strongly affects the propagation of perturbations in the quark gluon plasma.
This conclusion was obtained with the relativistic Navier-Stokes formalism and  it would be interesting to check if it remains valid in
other relativistic theories of viscosity.

\section{Appendix}

The derivation of Eq. (\ref{bwqcdTfin_va}) is quite similar to derivation of  Eq. (\ref{bwqcdTfinest}) and here we would like to
add some details. The expansion of Eqs. (\ref{rnsagain}) and (\ref{relcontss}) in powers of $\sigma$ is straightforward. The calculation
is faster if we keep terms only up to $\sigma^2$ and neglect some higher order terms even before reaching the identities equivalent to
(\ref{sigma1}), (\ref{sigma2}) and (\ref{sigma3}). It is useful to remember that:
$$
\gamma^{n} \cong 1+(n/2) \times v^{2} \hspace{0.2cm} \propto  \hspace{0.2cm} 1 + \sigma^{2} +  \dots
$$
\begin{equation}
{\frac{\partial \gamma}{\partial t}}=\gamma^{3}v{\frac{\partial v}{\partial t}} \hspace{0.3cm} \propto \hspace{0.3cm} \sigma^{5/2}
\hspace{0.9cm} \textrm{and} \hspace{0.9cm} \vec{\nabla} \gamma=\gamma^{3}v\vec{\nabla} v \hspace{0.3cm}  \propto  \hspace{0.3cm} \sigma^{5/2}
\label{derivgamasigma}
\end{equation}
The viscous terms in (\ref{rnsagain}) are of the following order in $\sigma$:
\begin{equation}
\eta \, v^{i}\partial_{\mu}\partial^{\mu}\gamma \hspace{0.3cm} \propto  \hspace{0.3cm} \sigma^{4}
\label{vi1}
\end{equation}
\begin{equation}
\eta\, v^{i}\partial_{\mu}{\frac{\partial u^{\mu}}{\partial t}} \hspace{0.3cm} \propto  \hspace{0.3cm} \sigma^{3}
\label{vi2}
\end{equation}
\begin{equation}
\eta \, v^{i}\partial_{\mu}\bigg[\gamma\bigg({\frac{\partial}{\partial t}}+\vec{v} \cdot \vec{\nabla}\bigg)(\gamma u^{\mu}) \bigg]\bigg\}
\hspace{0.3cm} \propto  \hspace{0.3cm} \sigma^{3}
\label{vi3}
\end{equation}
\begin{equation}
v^{i}\bigg(\zeta-{\frac{2}{3}}\eta \bigg){\frac{\partial}{\partial t}}\bigg[{\frac{\partial \gamma}{\partial t}}+\vec{\nabla}\cdot (\gamma \vec{v}) \bigg]
\hspace{0.3cm} \propto  \hspace{0.3cm} \sigma^{3}
\label{vi4}
\end{equation}
\begin{equation}
+v^{i}\bigg(\zeta-{\frac{2}{3}}\eta \bigg)\partial_{\mu} \bigg\{ \gamma u^{\mu} \bigg[{\frac{\partial \gamma}{\partial t}}+\vec{\nabla}\cdot (\gamma \vec{v}) \bigg]\bigg\}
\hspace{0.3cm} \propto  \hspace{0.3cm} \sigma^{3}
\label{vi5}
\end{equation}
$$
\eta \, \partial_{\mu}\partial^{\mu}(\gamma v^{i}) \hspace{0.3cm} \propto  \hspace{0.3cm} \sigma^{2}
$$
$$
\eta \, \partial_{\mu}\partial^{i}u^{\mu} \hspace{0.3cm} \propto  \hspace{0.3cm} \sigma^{2}
$$
$$
\eta \, \partial_{\mu}\bigg[\gamma\bigg({\frac{\partial}{\partial t}}+\vec{v} \cdot \vec{\nabla}\bigg)(\gamma v^{i} u^{\mu}) \bigg]\bigg\} \hspace{0.3cm} \propto  \hspace{0.3cm} \sigma^{2}
$$
$$
\bigg(\zeta-{\frac{2}{3}}\eta \bigg)\partial^{i} \bigg[{\frac{\partial \gamma}{\partial t}}+\vec{\nabla}\cdot (\gamma \vec{v}) \bigg]
\hspace{0.3cm} \propto  \hspace{0.3cm} \sigma^{2}
$$
\begin{equation}
\bigg(\zeta-{\frac{2}{3}}\eta \bigg)\partial_{\mu} \bigg\{ \gamma v^{i} u^{\mu} \bigg[{\frac{\partial \gamma}{\partial t}}+\vec{\nabla}\cdot (\gamma \vec{v}) \bigg] \bigg\} \hspace{0.3cm} \propto  \hspace{0.3cm} \sigma^{3}
\label{vi10}
\end{equation}
Using (\ref{vi1}) to (\ref{vi10}) in (\ref{rnsagain}) and keeping only terms up to $\mathcal{O}(\sigma^{2})$ we obtain:
$$
(\varepsilon+p)\gamma^{2} \bigg({\frac{\partial}{\partial t}}+\vec{v} \cdot \vec{\nabla} \bigg) v^{i}+
v^{i}{\frac{\partial p}{\partial t}} - \partial^{i}p
$$
\begin{equation}
+\eta \, \bigg[ \gamma \, \bigg( {\frac{\partial^{2}v^{i}}{\partial t^{2}}}
-\vec{\nabla}^{2}v^{i} \bigg)+\gamma \, \partial^{i}(\vec{\nabla} \cdot \vec{v})
- \gamma^{3} \, {\frac{\partial^{2}v^{i}}{\partial t^{2}}}\bigg]
+\bigg(\zeta-{\frac{2}{3}}\eta \bigg)\gamma \, \partial^{i} (\vec{\nabla}\cdot \vec{v})=0
\label{nsorc}
\end{equation}
The $\gamma^{n}$ (with $n=1,2,3$) factors in (\ref{nsorc}) are also multiplying $v^{i}$ or its derivative, which
contributes with at least one power of $\sigma$.  Therefore we consider $\gamma^{n} \cong 1$ and (\ref{nsorc}) becomes:
\begin{equation}
{\frac{\partial \vec{v}}{\partial t}} +(\vec{v} \cdot \vec{\nabla}) \vec{v}=
-{\frac{1}{(\varepsilon+p)}}\bigg[\vec{\nabla} p + \vec{v}{\frac{\partial p}{\partial t}} \bigg] +
{\frac{1}{(\varepsilon+p)}}
\bigg[\eta \, {\vec{\nabla}}^{2}\vec{v}+
\bigg(\zeta+{\frac{1}{3}}\eta \bigg)
\vec{\nabla}(\vec{\nabla}\cdot \vec{v}) \bigg]
\label{nsorcend}
\end{equation}
The equation above is the simplified version of the relativistic Navier-Stokes equation and from
(\ref{nsorcend}) it is easier to derive (\ref{bwqcdTfin_v}).

Analogously we estimate the order (in $\sigma$) of  the viscous terms in (\ref{relcontss}):
\begin{equation}
{\frac{\eta}{T}}\bigg({\frac{\partial \gamma}{\partial t}}\bigg)^{2}
\propto  \hspace{0.3cm} \sigma^{5}
\label{vi11}
\end{equation}
\begin{equation}
{\frac{\eta}{T}}\bigg[\vec{\nabla} \gamma \cdot {\frac{\partial}{\partial t}}(\gamma\vec{v})
\bigg] \hspace{0.3cm} \sigma^{4}
\label{vi11a}
\end{equation}
\begin{equation}
{\frac{\eta}{T}}(\partial^{i}u^{j})\partial_{j}u_{i}
\propto  \hspace{0.3cm} \sigma^{3}
\label{vi12}
\end{equation}
\begin{equation}
{\frac{1}{T}}\bigg({\frac{2}{3}}\eta+\zeta \bigg) \, \bigg[ {\frac{\partial \gamma}{\partial t}}
+\gamma  \vec{\nabla} \cdot \vec{v} +\vec{\nabla}\gamma \cdot \vec{v} \bigg]^{2}
\propto  \hspace{0.3cm} \sigma^{3}
\label{vi13}
\end{equation}

\begin{acknowledgments}
The authors are grateful to R. Venugopalan, R.P.G. Andrade, S.B. Duarte, J. Noronha and M. Strickland for enlightening discussions.
This work was  partially financed by the Brazilian funding agencies CAPES, CNPq and FAPESP.
\end{acknowledgments}

\end{document}